\documentclass[onecolumn]{els-mrw-modified} 

\usepackage{amsmath,amssymb,amsfonts,makeidx,graphicx}
\usepackage{txfonts}
\usepackage{helvet} 

\newcommand{\vx}{\boldsymbol{x}}
\newcommand{\ve}{\boldsymbol{e}}
\newcommand{\eps}{\varepsilon}
\DeclareMathAlphabet{\bbold}{U}{bbold}{m}{n}
\newcommand{\id}{\ensuremath{\bbold{1}}}  
\newcommand{\ud}{\mathrm{d}}
\newcommand{\ui}{\mathrm{i}}
\newcommand{\ue}{\mathrm{e}} 

\newcommand{\Tr}{\operatorname{Tr}}
\renewcommand{\Re}{\operatorname{Re}}
\renewcommand{\Im}{\operatorname{Im}} 
\newcommand{\bq}{\boldsymbol{q}}
\newcommand{\bp}{\boldsymbol{p}} 
\newcommand{\byy}{\boldsymbol{y}}

\begin{document}

\chapter{Semiclassical  periodic-orbit theory for quantum spectra}\label{chap1}

\author[1]{Sebastian M\"uller}

\author[1]{Martin Sieber}

\address[1]{\orgname{University of Bristol}, \orgdiv{School of Mathematics}, \orgaddress{Fry Building, Woodland Road, BS8 1UG, United Kingdom}}

\articletag{Chapter Article tagline: Sept. 1, 2025}

\maketitle

\begin{glossary}[Keywords]
Quantum Chaos, Semiclassical Methods, Periodic-Orbit Theory, Trace Formula, Spectral Statistics

\end{glossary}

\begin{abstract}[Abstract]
Gutzwiller's trace formula has a central place in quantum chaos because it provides semiclassical approximations for quantum energy levels 
in classically chaotic systems by linking them to classical periodic orbits.  In this didactic article, we discuss a derivation of the trace formula 
starting from the Feynman path integral. We then describe how the trace formula is used to explain universal features in the distribution of the
quantum energy levels that are described by random matrix theory, and we give an overview of related work. 
\end{abstract}

\section{Introduction}

In quantum chaos, our aim is to study the quantum mechanical features of systems whose classical limit is chaotic. Universal quantum signatures of chaos have been found, for example, in the statistics of energy levels, in the behavior of wavefunctions, and in transport properties such as the conductance.  To understand how these properties are affected by classical chaos, it is vital to develop methods that can relate quantum mechanics to classical mechanics.
These methods apply in the semiclassical limit, i.e., if the classical actions of the system are much larger than $\hbar$, which is typically connected to large energies and system sizes much larger than the quantum mechanical  wavelength.

In this chapter, our focus will be on the quantum mechanical spectra.
In one dimension, the WKB method allows the semiclassical approximation of individual energy levels and wave functions. This approach can
be generalized to higher dimensional systems, in the form of the EBK approximation \cite{Einstein-1917,Brillouin-1926,Keller-1958}, in cases where the classical motion is integrable.
However, Einstein \cite{Einstein-1917} already recognized that these
methods cannot be generalized to non-integrable systems. For chaotic systems, the Gutzwiller trace formula \cite{Gutzwiller-1970,Gutzwiller-1971}
provides an alternative approach to the semiclassical approximation of quantum energy levels. 

By relating quantum energies to classical
periodic orbits the trace formula allows to determine energies from the knowledge of classical periodic orbits.
This relation between energy levels and periodic orbits is not one-to-one, and in 
practice the use of the trace formula as a numerical tool
is limited because the number of periodic orbits in chaotic systems increases exponentially with their length.

However, the trace formula is crucial for statistical approaches.
One of the
central discoveries in quantum chaos is the observation that statistical distributions of energy levels are universal and agree
with those of random matrix theory \cite{Bohigas-1984}. As will be discussed in detail in sections 3 and 4 of this chapter, the trace formula can be
applied to show that the universal spectral statistics are related to correlations between periodic orbits that are universal in typical chaotic systems.

This chapter is organized as follows. Section 2 gives a didactic account (including exercises) of semiclassical methods and the derivation of the trace formula. We have aimed to make this section suitable as a first exposure to semiclassics, and we recommend the textbooks \cite{Haake-2018,Stockmann-1999,Wimberger-2022,Brack-1997,Cvitanovic-2025,Gutzwiller-1990} for further details. To establish a connection between quantum and classical mechanics, we   define the propagator, a quantity that can be used to describe the time evolution of quantum mechanical systems. We then express this propagator in terms of classical trajectories, first as an exact relation  to classical trajectories that are not required to satisfy the laws of motion (the Feynman path integral). Afterwards, we  derive an approximation, the Van Vleck-Gutzwiller propagator, which involves only trajectories satisfying these laws of motion. Finally, we proceed from the time domain to the energy domain and hence obtain Gutzwiller's trace formula relating the energy levels of chaotic systems to classical periodic orbits.

We then move on to the semiclassical theory of spectral statistics. Complementary reviews of this topic can be found in the textbook \cite{Haake-2018} and in a shorter book chapter by the present authors \cite{Muller-2011}. Section 3 gives an introduction into key background including the definition of the relevant statistical quantities (the correlation function and the spectral form factor), random-matrix theory, and classical chaos. In the present volume, more details on this background are given in \cite{Kieburg-2026,Tomsovic-2026}. To illustrate the use of semiclassical methods, we derive the first two contributions to series expansions of the correlation function from pairs of periodic orbits. In the diagonal approximation, the leading contribution is obtained from pairs of orbits that are identical or mutually time-reversed \cite{Hannay-1984,Berry-1985}. The next-to-leading contribution is due to pairs where one orbit contains a self-crossing with a small angle and its partner narrowly avoids this crossing \cite{Sieber-2001}, a mechanism that is helpful to recast in phase-space language. 

In section 4, we then summarize the essential ideas needed to semiclassically resolve the full expansion \cite{Muller-2009,Haake-2018}, including a refinement of the semiclassical approach and links to field theoretical methods.

We conclude with an overview of related work in section 5.

\section{Semiclassics and Gutzwiller's trace formula}

\subsection{The propagator}

Following the program laid out above, we show how the propagator can be used to describe quantum-mechanical time evolution, and how it can be related to classical paths first in exact and then in approximate form.

We consider conservative quantum systems with Hamiltonian
\begin{equation} \hat{H} = \frac{\hat{\bp}^2}{2m} + V(\hat{\bq}) .\end{equation}
For these systems, solutions $ | \psi(t) \rangle $ of the time-dependent Schr\"odinger equation\begin{equation}  \ui \hbar \frac{\partial}{\partial t} \, | \psi(t) \rangle = \hat{H}  \, | \psi(t) \rangle , \qquad
\text{with}  \qquad  | \psi(0) \rangle =  | \psi_0 \rangle,\end{equation}
may be written as 
\begin{equation} | \psi(t) \rangle = \hat{U}(t) \, | \psi_0 \rangle \end{equation}
where $\hat U(t)$ is the {\bf time-revolution operator} given by
\begin{equation} \label{timeev}\hat{U}(t) = \exp \left( - \frac{\ui}{\hbar} \hat{H} t \right) .\end{equation}  
If we now express $ | \psi(t) \rangle $ in position representation,
the time evolution (\ref{timeev})
can be expressed as
\begin{equation}\psi(\bq,t) = \langle \bq | \psi (t) \rangle = \int_{-\infty}^\infty \! \ud^f q_0 \, \langle \bq | \hat{U}(t) | \bq_0 \rangle \, \langle \bq_0 | \psi_0 \rangle 
= \int_{-\infty}^\infty \! \ud^f q_0 \, K(\bq,\bq_0,t) \, \psi_0(\bq_0) ,\end{equation}
where $f$ is the number of degrees of freedom and we have introduced the {\bf propagator} $K(\bq,\bq_0,t)$   
\begin{equation}K(\bq,\bq_0,t) = \langle \bq | \hat{U}(t) | \bq_0 \rangle . \end{equation}
In terms of eigenstates of the Hamiltonian $\hat{H} | \psi_n \rangle = E_n | \psi_n \rangle$ we have 
\begin{equation} \label{propen}K(\bq,\bq_0,t) = \sum_n \langle \bq | \hat{U}(t) | \psi_n \rangle \langle \psi_n | \bq_0 \rangle = \sum_n \psi_n(\bq)  \, \ue^{-\frac{\ui}{\hbar} E_n t } \, \psi_n^*(\bq_0), \end{equation}
Hence, the propagator contains all the knowledge about the eigenvalues $E_n$ and eigenstates $|\psi_n \rangle$ of the Hamiltonian. 

\begin{BoxTypeA}

\noindent 
{\bf Exercise:} Show that the propagator of the one-dimensional free particle is given by
\begin{equation} K(q,q_0,t) = \left\langle q \left| \exp \left( - \frac{\ui}{\hbar} \frac{\hat{p}^2}{2m} t \right) \right| q_0 \right\rangle =
\sqrt{\frac{m}{2 \pi \ui \hbar t}} \exp \left( \frac{\ui}{\hbar} \, \frac{m \, (q - q_0)^2}{2 t} \right) .\end{equation}
\bigskip
Hint: Use the completeness of the momentum eigenstates
\begin{equation} \hat{p} | p \rangle = p | p \rangle , \qquad \langle p | p' \rangle = \delta(p - p') , \qquad \int_{-\infty}^\infty \! \ud p \; | p \rangle \langle p | = \id .\end{equation}
the inner product
\begin{equation} \langle q | p \rangle = \frac{1}{\sqrt{2 \pi \ui \hbar}} \ue^{ \frac{\ui}{\hbar} p q } = {\langle p | q \rangle}^* ,\end{equation}
and the shifted Gaussian integral
\begin{equation} \int_{-\infty}^\infty \ud x \, \exp(-a x^2 + b x) = \sqrt{\frac{\pi}{a}} \exp \left\{ \frac{b^2}{4 a} \right\} ,\end{equation}
where $\Re a > 0$, or $\Re a, \, \Re b = 0$, $\Im a \neq 0$. Here and in the following
$ \sqrt{\ui} = \ue^{\ui \frac{\pi}{4}}, \; \sqrt{-\ui} = \ue^{-\ui \frac{\pi}{4}} .$
Show further that the propagator of the $f$-dimensional free particle is the product of $f$ one-dimensional propagators.

\end{BoxTypeA}

\subsubsection*{Lagrangian mechanics}

Before we proceed to relate the propagator to classical paths, we make a brief excursion into Lagrangian mechanics, primarily because we will build on some results presented as exercises.

In the Lagrangian formulation of mechanics, the Lagrangian is defined as the difference between kinetic and potential energy
\begin{equation} L(\bq,\dot{\bq}) = T - V = \frac{m}{2} \dot{\bq}^2 - V(\bq) , \end{equation}
where $\bq=(q_1,\ldots,q_f)$ in $f$ dimensions. The equations of motion are given by the Euler-Lagrange equations
\begin{equation} \frac{d}{d t} \frac{\partial L}{\partial \dot{\bq}} - \frac{\partial L}{\partial \bq} = 0,  \end{equation}
which can be shown to agree  with Newton's second law. It is a central property of the Lagrangian formalism that the equations of motion can be obtained from an action principle. The action is defined for trajectories $\bq(t)$ that run from $\bq_a$ at time $t_a$ to $\bq_b$ at time $t_b$,
\begin{equation} R[\bq(t)] = \int_{t_a}^{t_b}  L(\bq,\dot{\bq}) \, \ud t .\end{equation}
The classical trajectories are those that make the functional $R[\bq]$ stationary with respect to infinitesimal variations $\bq(t) \to \bq(t) + \delta \bq(t)$
that leave the end points invariant, $\delta \bq(t_a) = \delta \bq(t_b) = 0$.
The variation of $R[\bq]$ is given by 
\begin{align} 
0 = \delta R[\bq] = R[\bq+\delta \bq] - R[\bq] & =  \int_{t_a}^{t_b}  \left[ \frac{\partial L}{\partial \dot{\bq}} \delta \dot{\bq} + \frac{\partial L}{\partial \bq} \delta \bq \right]  \, \ud t   = \left. \frac{\partial L}{\partial \dot{\bq}} \delta \bq \right|^{t_b}_{t_a} - 
 \int_{t_a}^{t_b}  \left[ \frac{d}{d t} \frac{\partial L}{\partial \dot{\bq}} - \frac{\partial L}{\partial \bq} \right] \, \delta \bq \, \ud t , \notag
\end{align} 
where we integrated by parts. The first term in the last line vanishes due to $\delta \bq(t_a) =\delta \bq(t_b)=0$.
Hence, the variation of $R[\bq]$ vanishes if $\bq(t)$ satisfies the Euler-Lagrange equations.
Crucially, there can be several solutions $\bq(t)$ that connect $\bq_a$ at time $t_a$ to $\bq_b$ at time $t_b$.
In the following, we will mostly set $\bq_a=\bq_0$, $\bq_b=\bq$, $t_a=0$, $t_b=t$ for simplicity.

\begin{BoxTypeA}

\noindent
{\bf Exercise:} For a given classical trajectory, we can regard the action as a function $R(\bq_b,t_b,\bq_a,t_a)$ of initial and final positions and times, also known as the principal function. Show that Hamilton's principal function for a classical trajectory satisfies the relations
\begin{equation} \label{Rderiv}\frac{\partial R}{\partial \bq_b} = \bp_b , \qquad \frac{\partial R}{\partial \bq_a} = - \bp_a , \qquad \frac{\partial R}{\partial t_b} = -E ,\end{equation}
Hint: There are different ways to prove this. For example,  you can use the expression for $\delta R[\bq]$ to obtain the first two relations. 
The third relation is more tricky. One way is to fix a classical trajectory and follow it longer in time
\begin{equation}\int_{t_a}^{t_b+\delta t_b}  L(\bq,\dot{\bq}) \, \ud t =   R (\bq_b + \dot{\bq}_b\delta t_b,\;  t_b+\delta t_b,\;\bq_a,\;t_a) ,\end{equation}
Then the required result can be extracted from a first order expansion of this expression.

\end{BoxTypeA}

\begin{BoxTypeA}

\noindent
{\bf Exercise:} Show that the propagator of the free particle can be written in the form
\begin{equation} K(q_b,t_b,q_a,t_a) = \label{free}\sqrt{\frac{1}{2 \pi \hbar} \left| \frac{\partial ^2 R}{\partial q_b \partial q_a} \right| }  \exp \left( \frac{\ui}{\hbar} \, R(q_b,t_b,q_a,t_a) - \ui \frac{\pi}{4} \right) ,\end{equation}
where $R(q_b,t_b,q_a,t_a)$ is Hamilton's principal function for the trajectory of the free particle from $q_a$ at time $t_a$ to $q_b$ at time $t_b$.

\end{BoxTypeA}

\subsection{Feynman path integral}

\begin{figure}[t]
\centering
\includegraphics[width=.5\textwidth]{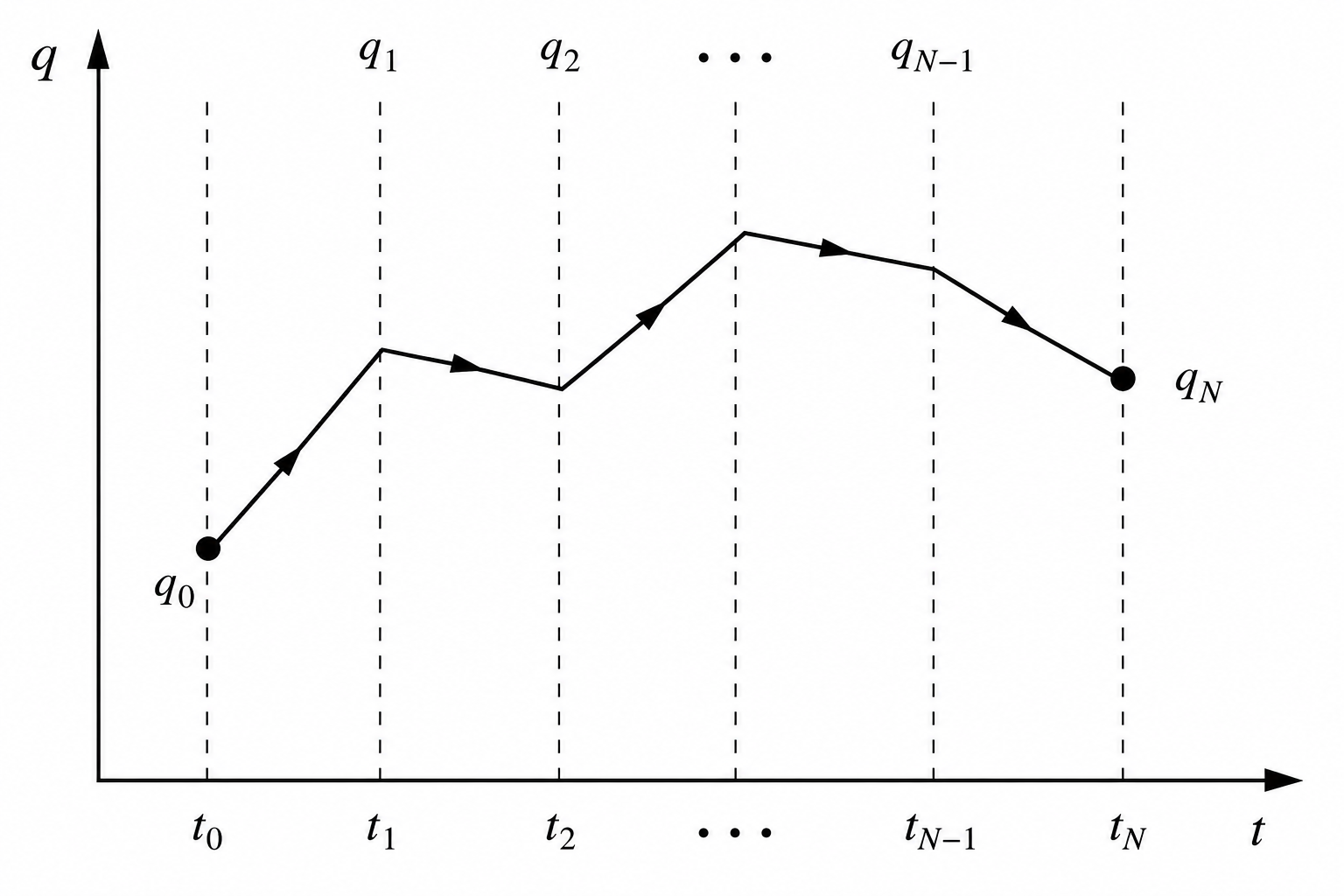}
\caption{The Feynman path integral extends over broken line paths like the one in this figure.}
\label{bl_fig1}
\end{figure}

We are now ready to express the propagator as an integral over paths \cite{Feynman-1948,Haake-2018,Altland-2023}. We restrict ourselves to the one-dimensional case, which illustrates the main features, but the result holds more generally. A key ingredient is that the time evolution can be split into parts
\begin{equation} \hat{U}(t) = \ue^{- \ui \hat{H} t / \hbar} = \ue^{- \ui \hat{H} t_1 / \hbar} \, \ue^{- \ui \hat{H} t_2 / \hbar}  = \hat{U}(t_2) \hat{U}(t_1) ,\end{equation}
where $t=t_1 + t_2$. For the propagator, this corresponds to the composition property
\begin{equation} \label{compose}K(q_c, q_a, t) =   \langle q_c | \hat{U}(t) | q_a \rangle =  \langle q_c |  \hat{U}(t_2) \hat{U}(t_1) | q_a \rangle =
\int_{-\infty}^\infty \! \ud q_b \, K(q_c, q_b, t_2)  \, K(q_b, q_a, t_1) .\end{equation}
The idea of the Feynman path integral is to split the time evolution into a large number $N$ of parts $\tau=\frac{t}{N}$ and take
the limit $N \to \infty$. Thus, we consider
\begin{equation} \hat{U} (t) = \left[ {\hat{U}} \left(\tau\right) \right]^N \end{equation}
This brings in time evolutions over short times, 
which we can afterwards split into factors associated with the kinetic and potential energy as\begin{equation}\hat{U} \left(\tau\right) = \exp \left(- \frac{\ui \tau \hat{H}}{\hbar  } \right)   = \exp \left(- \frac{\ui \tau (\hat{T}+\hat{V})}{\hbar  } \right)   = \exp \left(- \frac{\ui \tau \hat{T}}{\hbar  } \right)   \exp \left(- \frac{\ui \tau \hat{V}}{\hbar  } \right) 
 + {\cal O}\left(\tau^2 \right) .\end{equation}
The error term follows from a Taylor expansion and can be neglected when taking the limit
$N \to \infty$. We now obtain
\begin{align}
K(q,q_0,t) & = \langle q | [  \ue^{- \frac{\ui\tau}{\hbar} (\hat{T} + \hat{V})} ]^N | q_0 \rangle \notag \\
& = \lim_{N \to \infty} \langle q | [ \ue^{-  \frac{\ui\tau}{\hbar} \hat{T}} \ue^{- \frac{\ui\tau}{\hbar}{\hat V}}  ]^N | q_0 \rangle \notag \\
& = \lim_{N \to \infty} \int_{-\infty}^\infty \! \ud q_1 \ldots \ud q_{N-1} \prod_{j=0}^{N-1} \langle q_{j+1} | 
\ue^{- \frac{\ui\tau}{\hbar} \hat{T}} \ue^{- \frac{\ui\tau}{\hbar} \hat{V}}  | q_j \rangle \notag \\ \notag
& = \lim_{N \to \infty} \int_{-\infty}^\infty \! \ud q_1 \ldots \ud q_{N-1} \prod_{j=0}^{N-1} \langle q_{j+1} | 
\ue^{- \frac{\ui\tau}{\hbar}\hat{T}} | q_j \rangle \,  \ue^{- \frac{\ui\tau}{\hbar} V(q_j) } \notag 
\end{align} 
where $q_N=q$. If we now use the propagator of the free particle (\ref{free}) for $\langle q_{j+1} | 
\ue^{- \frac{\ui\tau}{\hbar}\hat{T}} | q_j \rangle$, we get
\begin{equation} \label{pathintdisc} K(q,q_0,t)  =\lim_{N \to \infty} \left(\frac{m}{2 \pi \ui \hbar \tau} \right)^{N/2}\int_{-\infty}^\infty \! \ud q_1 \ldots \ud q_{N-1} 
\exp \left( \frac{\ui\tau}{\hbar}  \sum_{j=0}^{N-1} \left[ \frac{m}{2} \, \frac{(q_{j+1} - q_j )^2}{\tau^2} - V(q_j) \right] \right). \end{equation}
The exponent involves the action
\begin{equation}   \sum_{j=0}^{N-1} \left[ \frac{m}{2} \, \frac{(q_{j+1} - q_j )^2}{\tau^2} - V(q_j) \right] \tau\;\substack{\longrightarrow \\ N \to \infty} 
\int_0^t \ud t' \left[ \frac{m}{2} \dot{q}^2(t') - V(q(t')) \right]  = R(q,q_0,t) . \end{equation}
It is natural to interpret the integration variables $q_j$ in (\ref{pathintdisc}) as positions after time $j\tau$, with fixed initial position $q_0$ and final position $q_N=q$.
The integral is thus taken over broken-line paths as shown in Fig.\ref{bl_fig1}. 
If we formally take the limit $N\to\infty$, we obtain
an integral over all paths that go from $q_0$
to $q$ in time $t$, written symbolically as
\begin{equation} \label{pathint} K(q,q_0,t) = \int_{\stackrel{q(t)=q}{\scriptscriptstyle q(t_0) = q_0}} \! {\cal D}[q] \, \exp \left\{ \frac{\ui}{\hbar} R[q] \right\} . \end{equation}
We note that the paths here are not required to satisfy the classical equations of motion. Equations (\ref{pathintdisc}) and (\ref{pathint}) represent the {\bf Feynman path integral} in discrete and continuous form.

In the following, we will approximate the integral in the semiclassical limit, using
 a stationary-phase approximation. The result will then express the propagator as the sum over all trajectories that go from $q_0$ to $q$ in time $t$ and satisfy the classical equations of motion. As a preparation, we include a short section on the stationary-phase approximation.

\subsection{Stationary-phase approximation}

The stationary-phase approximation provides the leading order approximation as $\lambda \to \infty$ for the highly oscillatory integral of the form
\begin{equation} I  = \int_a^b \! \ud y \; A(y) \, \ue^{\ui \, \lambda \, \phi(y)}\end{equation}
We assume that $\phi$ and $A$ are  smooth,  that $\phi$ has one non-degenerate stationary point $\phi'(y_0)=0$, $\phi''(y_0) \neq 0$, with $a< y_0 < b$, and that $A(y_0)\neq 0$. Then the leading order approximation
in the limit $\lambda \to \infty$ is obtained by expanding the phase up to second order about the stationary point
\begin{equation}\phi(y) \approx \phi(y_0) + \frac{1}{2} \phi''(y_0) (y - y_0)^2\end{equation} 
and using the Fresnel (complex Gaussian) integral
	\begin{equation}  \int_{-\infty}^\infty \! \ud y \; \ue^{\ui c y^2}  
	=\sqrt{\frac{\ui\pi}{c}}=	\sqrt{\frac{\pi}{|c|}}\ue^{\ui\frac{\pi}{4}{\rm sgn}\, c} .\end{equation} 
Here, the final expression fixes the branch of the square root left implicit in the preceding term. The factor $c$ is  assumed to be real and different from 0, with ${\rm sgn}\,c=1$ for
positive $c$ and ${\rm sgn}\,c=-1$ for
negative $c$. 	
	The final result is then
\begin{equation} I \label{1dstat} \approx \int_{-\infty}^\infty \! \ud y \, A(y_0) \, \ue^{\ui \lambda\phi(y_0) + \frac{\ui \lambda}{2} |\phi''(y_0)| \, (y - y_0)^2}
=  A(y_0) \, \sqrt{\frac{2 \pi}{\lambda |\phi''(y_0)|}} \, \, \ue^{\ui \lambda \phi(y_0)+\ui\frac{\pi}{4}{\rm sgn}\,\phi''(y_0)}
\qquad \text{as} \qquad \lambda \to \infty.\end{equation}
We note the following points:
\begin{itemize}
\item The main contribution to the integral comes from a region of order $\Delta y \propto \lambda^{-1/2}$.
\item If there are several stationary points in the interval $(a,b)$ the contributions are added.
\item For a stationary point at one of the endpoints, $a$ or $b$, we get half the contribution.
\item If there is no stationary point in the interval $[a,b]$ then the leading order approximation
(of order $1/\lambda$) comes from the end points. It can be obtained by integration by parts.
\end{itemize}
For an {\bf arbitrary number of dimensions $f$}, we instead consider the integral
\begin{equation} I  = \int \! \ud^f y \; A(\byy) \, \exp \left( \ui \, \lambda \, \phi(\byy) \right). \end{equation}
In this case, the contribution of a stationary point $\nabla \phi(\byy_0) =0$   is given by
\begin{equation}   A(\byy_0) \left( \frac{2 \pi}{\lambda} \right)^{f/2} \left| \det \frac{\partial^2 \phi}{\partial y_i \partial y_j} (\byy_0) \right|^{-1/2} \ue^{\ui \lambda \phi(\byy_0) + \ui  \nu\pi/4} , \end{equation}
where $\nu$ is the number of positive eigenvalues minus the number of negative eigenvalues of the matrix of second order derivatives, and it is assumed that this
matrix does not have a vanishing determinant at $\byy_0$. One can show that this means that the stationary points must be {\bf isolated}.


\subsection{Semiclassical propagator}

We now {\bf apply the stationary-phase approximation to the path integral}. We do this in the semiclassical limit, where $\hbar$ is taken to 0 or, more precisely,  assumed to be much smaller than all relevant classical actions. Conceptually, the answer is clear: We should obtain a sum over all trajectories for which the phase, i.e. in this case the action, becomes stationary with respect to small variations. Due to Hamilton's principle, these are just the classical paths satisfying the equations of motion. 

The stationary-phase approximation can be carried out starting from the discrete-time version of the path integral 
\begin{equation} \left(  \frac{m}{2 \pi \ui \hbar \tau} \right)^{N/2}\int_{-\infty}^\infty \! \ud q_1 \ldots \ud q_{N-1} 
\exp \left( \frac{\ui}{\hbar}  \sum_{j=0}^{N-1} \left[ \frac{m}{2} \, \frac{(q_{j+1} - q_j )^2}{\tau^2} - V(q_j) \right]\tau \right) .\end{equation}
Denoting the exponent by $\ui \phi / \hbar$ we obtain the stationary points from
\begin{equation} 0 = \frac{\partial \phi}{\partial q_j} = \left[ - m \frac{(q_{j+1} - q_j) - (q_j - q_{j-1})}{\tau^2}
- \frac{\partial V(q_j)}{\partial q_j} \right]\tau  \qquad j=1, \ldots , N-1 .\end{equation}
This is a discretized form of Newton's equations, and in the limit $\tau\to 0$ we obtain 
\begin{equation} - m \ddot{q} - \frac{\partial V}{\partial q} = 0 .\end{equation}
Hence, the semiclassical limit of the propagator is indeed given by a sum over all classical paths from $q_0$ to $q$ in time $t$
\begin{equation} K(q,q_0,t) \approx \sum_\gamma D_\gamma \, \ue^{\frac{\ui}{\hbar} R_\gamma} . \end{equation}

The prefactor $D_{\gamma}$ can be obtained by performing the remaining steps of the stationary-phase approximation in $N-1$ dimensions \cite{Morette-1951,Haake-2018}.
However, this calculation is tricky. 
Alternative derivations \cite{VanVleck-1928, Stockmann-1999} can be based on the Schr\"odinger equation or on the {\bf composition property} (see (\ref{compose}))
\begin{equation} K(q_c, q_a, t) = \int_{-\infty}^\infty \! \ud q_b \, K(q_c, q_b, t_2)  \, K(q_b, q_a, t_1) \end{equation}
where $t = t_2 + t_1$.
Here, we will use the composition property and focus on the contribution of one trajectory, intuitively labeled by subscripts $ba$, to $K(q_b, q_a, t_1)$ and the contribution of another trajectory labeled by $cb$ to $K(q_c, q_b, t_2)$. 
If we evaluate this integral in a stationary-phase approximation, 
now in only one dimension,
the stationary-phase condition is
\begin{equation} \frac{\partial R_{cb}}{\partial q_b} + \frac{\partial R_{ba}}{\partial q_b}=0\,.	
\end{equation}
Using (\ref{Rderiv}), the first summand is minus the initial momentum of $cb$ and the second summand is plus the final momentum of $ba$.
This means that the two partial trajectories join smoothly and form
parts of a classical trajectory $ca$ from $q_a$ to $q_c$ in time $t$.
This combined trajectory has the action
\begin{equation} R_{ca} = R_{cb} + R_{ba} .\end{equation} 
Completing the stationary phase approximation results in
\begin{equation}\label{Dcont} D_{ca} \, \ue^{\frac{\ui}{\hbar} R_{ca}} = D_{cb} \, \ue^{\frac{\ui}{\hbar} R_{cb}}  D_{ba} \, \ue^{\frac{\ui}{\hbar} R_{ba}} \,
\sqrt{\frac{2 \pi \ui \hbar}{\left[ \frac{\partial^2 R_{cb}}{\partial q_b^2} + \frac{\partial^2 R_{ba}}{\partial q_b^2} \right]_{q_b = \bar{q}_b}}}   \end{equation}
where a classical calculation shows that
\begin{equation} \left[ \frac{\partial^2 R_{cb}}{\partial q_b^2} + \frac{\partial^2 R_{ba}}{\partial q_b^2} \right]_{q_b = \bar{q}_b} \, 
= - \frac{ \frac{\partial^2 R_{cb}}{\partial q_c \partial q_b} \, \frac{\partial^2 R_{ba}}{\partial q_b \partial q_a} }{ \frac{\partial^2 R_{ca}}{\partial q_c \partial q_a} } .\end{equation}
Eq. (\ref{Dcont}) thus simplifies to 
\begin{equation} D_{ca} = D_{cb} \, D_{ba} \, \sqrt{\frac{2 \pi \hbar}{\ui} \; \frac{\frac{\partial^2 R_{ca}}{\partial q_c \partial q_a} } { \frac{\partial^2 R_{cb}}{\partial q_c \partial q_b} 
\, \frac{\partial^2 R_{ba}}{\partial q_b \partial q_a}} }  \end{equation}
which has solutions of the type
\begin{equation}D_{ca} =\sqrt{\frac{1}{2 \pi \hbar} \, \left| \frac{\partial^2 R_{ca}}{\partial q_c \partial q_a}  \right| } \; \tilde{D}_{ca} . \end{equation}

The remaining question is: {\bf What is the phase factor $\tilde{D}_{ca}$?} The answer depends on the signs of the mixed partial derivatives. We have
\begin{equation} \frac{\partial^2 R_{ba}}{\partial q_b \partial q_a} = - \left. \frac{\partial p_a}{\partial q_b} \right|_{q_a} \qquad \text{or} \qquad 
\left[ \frac{\partial^2 R_{ba}}{\partial q_b \partial q_a} \right]^{-1} = - \left. \frac{\partial q_b}{\partial p_a} \right|_{q_a} \end{equation}
The last term involves the change $- \delta q_b$ in the final position $q_b$ of a trajectory that starts at $q_a$
with momentum $p_a + \delta p_a$. The mixed partial derivative changes sign when this quantity is zero.
\medskip
We need to consider trajectories in the neighborhood of a classical trajectory which satisfies Newton's second law
\begin{equation} m \frac{d^2}{d t^2} q= -V'(q) .\end{equation}
We replace $q(t)$ by $q(t) + \delta q (t)$ and obtain a differential equation for $\delta q(t)$
\begin{equation} m \frac{d^2}{d t^2} \delta q= - V''(q(t)) \, \delta q . \end{equation}
This is the {\bf Jacobi equation}. It is a homogeneous second order linear equation for $\delta q(t)$.
According to the {\bf Sturm separation theorem} \cite{Sturm-1836}, the zeros of any two linearly independent solutions of this equation are 
alternating and simple. 

Let $\nu_{ba}$ denote the number of zeros of a solution of the Jacobi equation of length $t_1$ that starts with $\delta q(0)=0$, $\delta \dot{q} (0) \neq 0$. Then 
\begin{equation}\left(  \frac{\partial^2 R_{ba}}{\partial q_b \partial q_a} \right) = \left|  \frac{\partial^2 R_{ba}}{\partial q_b \partial q_a} \right| \,  (-1)^{1+\nu_{ba}} .\end{equation}
There is an additional $(-1)$ because for small times the mixed partial derivative approaches that of a free particle and is negative. The $ac$ and $bc$ cases
are defined correspondingly, e.g.
\begin{equation}D_{ca} = \sqrt{\frac{1}{2 \pi \hbar} \, \left| \frac{\partial^2 R_{ca}}{\partial q_c \partial q_a}  \right| } \; \tilde{D}_{ca} \end{equation}
We then obtain
\begin{equation} \tilde{D}_{ca}  = \tilde{D}_{cb}  \, \tilde{D}_{ba}  \, \sqrt{\ui \, (-1)^{\nu_{ca} - \nu_{ba} - \nu_{cb} } } \, .\end{equation}
It now follows from the Sturm separation theorem that there are only two possibilities. 
\begin{equation}
\nu_{ca} - \nu_{ba} - \nu_{cb} = \begin{cases} 0 \\ 1 \end{cases} .
\end{equation}
For both cases, we find 
\begin{align}
\tilde{D}_{ca}  & = \tilde{D}_{cb}  \, \tilde{D}_{ba}  \, \sqrt{\ui \, (-1)^{\nu_{ca} - \nu_{ba} - \nu_{cb} } } \notag \\[2mm] \notag
& = \tilde{D}_{cb}  \, \tilde{D}_{ba} \; \exp \left( \ui \frac{\pi}{4} - \ui  ( \nu_{ca} - \nu_{ba} - \nu_{cb} ) \frac{\pi}{2}\right) .  
\end{align} 
This leads to our final result
\begin{equation} D_{ca} = \sqrt{\frac{1}{2 \pi \hbar} \, \left| \frac{\partial^2 R_{ca}}{\partial q_c \partial q_a}  \right| } \; \exp \left( - \ui \frac{\pi}{4} - \ui \nu_{ca} \frac{\pi}{2}  \right) . \end{equation}
Strictly speaking, there could be an additional factor $d_{ca}$ with $d_{ca} = d_{cb} \, d_{ba}$, but this factor is one. This can be seen, for example, in cases 
where the propagator is exact.
\bigskip

In summary, we thus obtain the following result for the semiclassical propagator
\begin{equation} \label{Ksc1} K(q,q_0,t) \approx \sum_\gamma \sqrt{\frac{1}{2 \pi \hbar} \, \left| \frac{\partial^2 R_\gamma}{\partial q \partial q_0}  \right| } 
\; \exp \left( \frac{\ui}{\hbar} R_\gamma(q,q_0,t) - \ui \frac{\pi}{4} - \ui \nu_\gamma  \frac{\pi}{2} \right) , \end{equation}
where the {\bf Morse index} $\nu_\gamma$ counts the number of times the prefactor has diverged, i.e., its inverse has gone through zero, along the trajectory.
The semiclassical propagator was first derived by Van Vleck \cite{VanVleck-1928} who still missed the Morse index. Morette \cite{Morette-1951} gave a derivation within the path integral formalism and Gutzwiller \cite{Gutzwiller-1967} realized the relevance of the Morse index. In the quantum chaos community, the semiclassical propagator is usually referred to as the 
{\bf  Van Vleck propagator} or the {\bf  Van Vleck-Gutzwiller propagator}.

\begin{BoxTypeA}

\noindent
{\bf Exercise:} For the harmonic oscillator with potential
\begin{equation} V(q) = \frac{m}{2} \omega^2 q^2 ,\end{equation}
the semiclassical approximation is identical to the exact result. Determine the propagator of the harmonic oscillator, valid for all times $t>0$. 
This formula is known as Feynman-Souriau formula.

\end{BoxTypeA}

In $f$ dimensions the semiclassical propagator turns into
\begin{equation}K(\bq,\bq_0,t) \approx \sum_\gamma \frac{1}{(2 \pi \hbar)^{f/2}} \sqrt{ \left| \det \left( \frac{\partial^2 R_\gamma}{\partial \bq \, \partial \bq_0} \right) \right| }
\exp \left\{  \frac{\ui}{\hbar} R_\gamma(\bq,\bq_0,t) - \ui f \frac{\pi}{4} - \ui \nu_\gamma \frac{\pi}{2} \right\}\end{equation} 
where the sum runs again over all classical trajectories $\gamma$ that go from $\bq_a$ to $\bq_b$ in time $t$. 
To specify the {\bf Morse index} $\nu_\gamma$ in the general case, we consider the matrix
\begin{equation}  A = \left( \frac{\partial^2 R}{\partial \bq \, \partial \bq_0} \right)^{-1}  = - \frac{\partial \bq}{\partial \bp_0} .\end{equation} 
The index $\nu_\gamma$ increases by one for every reduction of the rank of $A$ by one if one follows the trajectory over time $t$. 
This typically occurs at {\bf conjugate points}. At these points neighboring trajectories that start at the same initial position with infinitesimally 
different initial momentum intersect the original trajectory again. Conjugate points occur, for example, at {\bf caustics}. At conjugate points, the semiclassical
propagator diverges. In order to obtain an accurate approximation of the propagator at the conjugate points themselves, one has to replace
the stationary phase approximation by a uniform approximation that takes account of two or more close stationary points. 


Note that the form of the semiclassical propagator is familiar from {\bf unitary transformations}. Consider a classical canonical transformation from 
coordinates $(q,p)$ to coordinates $(Q,P)$. In quantum mechanics, the change of basis implied by this transformation is expressed by unitary transformations. Using the inner products of position and momentum eigenfunctions   
\begin{equation} \langle q | p \rangle = \frac{1}{\sqrt{2 \pi \hbar}} \ue^{ \frac{\ui}{\hbar} p q } = {\langle p | q \rangle}^* , \qquad \qquad
 \langle Q | P \rangle = \frac{1}{\sqrt{2 \pi \hbar}} \ue^{ \frac{\ui}{\hbar} P Q } =  {\langle P | Q \rangle}^* ,\end{equation}
the corresponding transition amplitudes between coordinates on either side of the canonical transformation are given by \cite{Dirac-1958,Miller-1974}
\begin{alignat}{2}
&   \langle q | Q \rangle = \left[ \frac{-1}{2 \pi \ui \hbar} \, \frac{\partial^2 F_1}{\partial q \partial Q} \right]^{1/2} \! \! \! \exp \left( \frac{\ui}{\hbar} F_1(q,Q) \right) , \quad 
&& \langle q | P \rangle = \left[ \frac{1}{2 \pi \ui \hbar} \, \frac{\partial^2 F_2}{\partial q \partial P} \right]^{1/2} \! \! \! \exp \left( \frac{\ui}{\hbar} F_2(q,P) \right) , \notag \\[2mm]  
&   \langle p | Q \rangle = \left[ \frac{1}{2 \pi \ui \hbar} \, \frac{\partial^2 F_3}{\partial p \partial Q} \right]^{1/2} \! \! \! \exp \left( \frac{\ui}{\hbar} F_3(p,Q) \right) , \quad 
&& \langle p | P \rangle = \left[ \frac{-1}{2 \pi \ui \hbar} \, \frac{\partial^2 F_4}{\partial p \partial P} \right]^{1/2} \! \! \! \exp \left( \frac{\ui}{\hbar} F_4(p,P) \right) ,
\end{alignat}
where the functions $F_i$ are the generating functions of the canonical transformation. 
If there are several stationary points, one has to  add their contributions.
The semiclassical propagator (\ref{Ksc1}) 
is of the same form, as the time evolution operator
is a unitary transformation. The corresponding classical canonical transformation
maps the coordinates $(q_0,p_0)$ at time $0$ to the coordinates $(q,p)$ at time $t$,
and its generating function is the principal function $R(q,q_0,t)$.

\subsection{Green function}

A first step to accessing energy levels semiclassically is to proceed from the time-dependent propagator to the energy-dependent Green function. The Green function is obtained as the {\bf Laplace transform} of the propagator
\begin{equation} \label{laplace}G(\bq,\bq_0,E^+) = \frac{1}{\ui \hbar} \lim_{\eta \to 0} \int_0^\infty \! \ud t \, K(\bq,\bq_0,t) \, \exp \left( \frac{\ui}{\hbar} E^+ \, t \right), \end{equation} 
leading to
\begin{equation} G(\bq,\bq_0,E^+) = \left\langle \bq \left| \frac{1}{E^+ - \hat{H}} \right| \bq_0 \right\rangle . \end{equation}
Here, the energy is taken with an infinitesimally small positive imaginary part $\eta$ as $E^+=E+\ui\eta$, to ensure the convergence of the integral.
The Green function is the kernel of the resolvent operator $(E^+ -\hat{H})^{-1}$ and satisfies the differential equation
\begin{equation} \left( E^+ -\hat H\right) G(\bq,\bq_0,E) = \delta(\bq - \bq_0) \, .\end{equation} 
Crucially, application of (\ref{laplace}) to the expression  (\ref{propen}) for the propagator yields the following representation in terms of eigenfunctions and eigenvalues of the Hamiltonian
\begin{equation} G(\bq,\bq_0,E^+) = \sum_n \frac{\psi_n(\bq) \, \psi^*_n(\bq_0)}{E^+ - E_n} , \end{equation}
which will eventually allow us to proceed from the Green function to the level density.
 
\begin{BoxTypeA}

\noindent
{\bf Exercise:} Consider the one-dimensional particle in a box, $q \in [0,L]$.
\begin{equation} - \frac{\hbar^2}{2m} \psi''(q) = E \psi(q) , \quad \psi(0) = 0 , \quad \psi(L) = 0 . \end{equation}
The Green function for a particle in a box satisfies
\begin{equation}  \left( E + \frac{\hbar^2}{2m} \frac{d^2}{dq^2} \right) G(q,q_0,E) = \delta(q - q_0) , \quad G(0,q_0,E)=G(L,q_0,E) = 0 ,\end{equation}
where we have dropped the imaginary part.
Show that it can be obtained from the general formula
\begin{equation} G(q,q_0,E) = \frac{2m}{\hbar^2} \, \frac{\psi_l(q_<) \, \psi_r(q_>)}{W(q_0)} . \end{equation}
Here $q_<$ and $q_>$ are the smaller and larger of $q$ and $q_0$, respectively. $\psi_l$ is a solution of the Schr\"odinger equation
that satisfies the boundary condition on the left-hand side, and $\psi_r$ correspondingly on the right-hand side. The Wronskian $W$ is defined as
\begin{equation} W(q)= \psi_l(q) \psi_r'(q) - \psi_l'(q) \psi_r(q) . \end{equation}
You need to show that the Green function satisfies both boundary conditions,
and that it satisfies the differential equation in the regions $q<q_0$ and $q>q_0$. The
delta function can be verified by integrating the differential equation from $q=q_0-\varepsilon$
to $q=q_0+\varepsilon$ and letting $\varepsilon$ go to zero. What is the result for the Green function?
\end{BoxTypeA}

If we Laplace transform the propagator of the free particle, we obtain the {\bf free Green function} in $f$ dimensions
\begin{equation} G_\mathrm{free}(\bq,\bq_0,E) = \frac{m}{2 \ui \hbar^2} \left( \frac{1}{2 \pi \hbar} \, \frac{\sqrt{2mE}}{|\bq - \bq_0|} \right)^{f/2-1} \, H_{f/2 - 1}^{(1)} \left( \frac{\sqrt{2mE}}{\hbar} | \bq - \bq_0 | \right) \, . \end{equation}
Here, $H_l^{(1)}(z) = J_l(z) + \ui N_l(z)$ denotes the Hankel function of the first kind.

Next, we want to obtain a {\bf semiclassical approximation for the general Green function}. For this purpose, we insert the semiclassical propagator
\begin{equation}K(\bq,\bq_0,t) \approx \sum_\gamma  \frac{1}{(2 \pi \hbar)^{f/2}} \, \sqrt{  \left| \det \frac{\partial^2 R_\gamma}{\partial \bq \, \partial \bq_0} \right| }
\exp \left\{  \frac{\ui}{\hbar} R_\gamma(\bq,\bq_0,t) - \ui \, f \frac{\pi}{4} - \ui \nu_\gamma \frac{\pi}{2} \right\}\end{equation}
into the transform
\begin{equation} G(\bq,\bq_0,E^+) = \frac{1}{\ui \hbar} \lim_{\eta \to 0} \int_0^\infty \! \ud t \, K(\bq,\bq_0,t) \, \exp \left( \frac{\ui}{\hbar} E^+ \, t \right).\end{equation} 
Dropping the imaginary part of $E^+$, we can write the phase in terms of the {\bf reduced action}
 \begin{equation}
S_\gamma=R_\gamma+Et
\end{equation}
and evaluate the integral over time in a stationary-phase approximation. This leads to the condition 
\begin{equation} \label{stattime}\frac{\partial R_\gamma(\bq,\bq_0,t)}{\partial t}  + E = 0  \qquad \text{or} \qquad - E_\gamma(t) + E = 0, \end{equation} 
which fixes the duration of the trajectory so that its energy coincides with $E$. 
Using the general formula for the one-dimensional stationary-phase approximation (\ref{1dstat}), we now obtain\begin{equation}G(\bq,\bq_0,E) \approx \sum_\gamma  \frac{2 \pi}{(2 \pi \ui \hbar)^{(f+1)/2}} \, \frac{ \sqrt{  \left| \det \frac{\partial^2 R_\gamma}{\partial \bq \, \partial \bq_0} \right|
}}{\sqrt{\left| \left. \frac{\partial^2 R_\gamma }{\partial t^2} \right. \right| }}
\exp \left\{  \frac{\ui}{\hbar} S_\gamma(\bq,\bq_0,E) - \ui \frac{\pi}{2} \xi_\gamma \right\}\end{equation}
where
\begin{equation} \xi_\gamma = \begin{cases} \nu_\gamma , \qquad & \frac{\partial^2 R}{\partial t^2} > 0 , \\ \nu_\gamma +1 , \qquad & \frac{\partial^2 R}{\partial t^2} < 0  \end{cases} \end{equation}
  can differ from the Morse index $\nu_\gamma$ by one. In the next step, we replace the actions $R_\gamma$ in the prefactor by the reduced action $S_\gamma$. As the calculation is quite technical, we only state the result, using a local coordinate
system in which one coordinate, indicated by a subscript $\parallel$, is taken along the trajectory and the remaining $(f-1)$ coordinates are perpendicular to the trajectory. We obtain  
\begin{equation}\label{Gsc}
G(\mathbf{q},\mathbf{q}_0,E)
\approx
\sum_{\gamma}
\frac{2\pi}{(2\pi \ui\hbar)^{(f+1)/2}}
\sqrt{
\left|
\frac{1}{\dot{q}_{\parallel}\dot{q}_{0\parallel}}
\det\left(
-\frac{\partial^2 S_\gamma}{\partial \mathbf{q}_{\perp}\,\partial \mathbf{q}_{0\perp}}
\right)
\right|
}
\exp\left[
\frac{\ui}{\hbar}S_\gamma(\mathbf{q},\mathbf{q}_{0},E)
-\ui\xi_\gamma\frac{\pi}{2}
\right],
\end{equation}
where the subscripts $\perp$ indicate $(f-1)$-dimensional vectors formed by perpendicular coordinates.

For later use, we also note the following {\bf derivatives of the reduced action} (taken as a function of $\bq$, $\bq_0$, and $E$, with the time fixed according to (\ref{stattime})
\begin{equation}\label{Sderiv} \frac{\partial S}{\partial \bq} = \bp , \qquad \frac{\partial S}{\partial \bq_0} = -\bp_0 , \qquad \frac{\partial S}{\partial E} = t . \end{equation}

\begin{BoxTypeA}

\noindent
{\bf Exercise:} Show that the Green function for a particle in a box, that you obtained in the previous exercise, can be written
as a sum over all trajectories from $q_0$ to $q$. The formula for a geometric series is helpful.
Convergence problems can be avoided by working with $E^+$ as defined above.
\end{BoxTypeA}

\subsubsection*{Stability matrix}

The prefactor in (\ref{Gsc}) can be expressed in terms of the {\bf stability matrix} $M$. This matrix describes the linearized 
motion in the neighborhood of the trajectory. It provides perpendicular deviations $\delta \bq_{\perp}$ and $\delta \bp_{\perp}$ at the
end point of a trajectory in terms of the corresponding deviations $\delta \bq_{0\perp}$ and $\delta \bp_{0\perp}$ at the starting point,
\begin{equation} \begin{pmatrix} \delta \bq_{\perp} \\ \delta \bp_{\perp} \end{pmatrix} = M \begin{pmatrix} \delta \bq_{0\perp} \\ \delta \bp_{0\perp} \end{pmatrix} . \end{equation}
$M$ is a symplectic matrix, which means that it satisfies $M^T J M = J$ where $J$ denotes the matrix {\footnotesize $J = \begin{pmatrix} 0 & I \\ -I & 0 \end{pmatrix} $}. In two dimensions, which we shall consider from now on, this is equivalent to $\det M = 1$. 
We can use (\ref{Sderiv}) to express
the stability matrix in terms of the reduced action   $S(\bq,\bq_0,E)$. With a single perpendicular direction, we have
\begin{equation} p_{\perp} = \frac{\partial S}{\partial q_{\perp}} , \qquad p_{0\perp} = - \frac{\partial S}{\partial q_{0\perp}} .\end{equation}
Linearization then gives
\begin{align}
\delta p_{\perp}  & =   \frac{\partial^2 S}{\partial q_{\perp}^2} \delta q_{\perp} +  \frac{\partial^2 S}{\partial q_{\perp} \partial q_{0\perp}} \delta q_{0\perp} , \notag \\  
\delta p_{0\perp} & = - \frac{\partial^2 S}{\partial q_{0\perp} \partial q_{\perp}} \delta q_{\perp} -  \frac{\partial^2 S}{\partial q_{0\perp}^2} \delta q_{0\perp} ,
\end{align}
and we can solve for $\delta q_{\perp}$ and $\delta p_{\perp}$ to obtain 
\begin{align}
\delta q_{\perp} & = \left( \frac{\partial^2 S}{\partial q_{\perp} \partial q_{0\perp}} \right)^{-1} \left[ -  \frac{\partial^2 S}{\partial q_{0\perp}^2} \delta q_{0\perp} - \delta p_{0\perp} \right], \notag \\
\delta p_{\perp} & = \left( \frac{\partial^2 S}{\partial q_{\perp} \partial q_{0\perp}} \right)^{-1} \left[ \left(\left( \frac{\partial^2 S}{\partial q_{\perp} \partial q_{0\perp}}\right)^2 
- \frac{\partial^2 S}{\partial q_{\perp}^2} \frac{\partial^2 S}{\partial q_{0\perp}^2} \right) \delta q_{0\perp} -  \frac{\partial^2 S}{\partial q_{\perp}^2}  \delta p_{0\perp} \right] 
\end{align} 
The elements of the stability matrix can now be read off and it is easy to check $\det M = 1$. With $M_{\gamma,12}=-\left(\frac{\partial^2 S_\gamma}{\partial q_{\perp} \partial q_{0\perp}} \right)^{-1}$, the semiclassical Green function simplifies to 
\begin{equation} \label{Gscsimp}G(\bq, \bq_0, E) \approx \sum_\gamma \frac{1}{\sqrt{ 2 \pi \hbar^3 \dot{q}_{\parallel} \dot{q}_{0\parallel} \, | M_{\gamma,12} | }} 
\exp \left( \frac{\ui}{\hbar} S_\gamma (\bq,\bq_0,E) - \ui \frac{\pi}{2} \xi_\gamma - \ui \frac{3 \pi}{4} \right) .
\end{equation} 
 Inverting the relation between the second derivatives and matrix elements, we can also show that
\begin{equation}\label{derivs}
\frac{\partial^2 S_\gamma}{\partial q_{\perp}^2}=\frac{M_{\gamma,22}}{M_{\gamma,12}},\quad\quad
\frac{\partial^2 S_\gamma}{\partial q_{0\perp}^2}=\frac{M_{\gamma,11}}{M_{\gamma,12}}.
\end{equation}

\subsection{Trace formula}

\subsubsection*{Density of states}

As we are specifically interested in the energy levels, we now consider the 
 {\bf density of states} (or level density) defined by the equation 
\begin{equation} 
\rho(E) = \sum_n \delta(E - E_n) .
\end{equation}
The density of states can be obtained from the Green function using the prescription
\begin{equation}
\label{trace}
	\rho(E)  = - \frac{1}{\pi} \lim_{\eta \to 0} \Im \int \ud^f q\, G(\bq,\bq,E + \ui \eta),
\end{equation}
which is easily checked using 
by writing the claimed formula as
\begin{align}  
  - \frac{1}{\pi} \lim_{\eta \to 0} \Im \int \! \ud^f q \sum_n \frac{\psi_n(\bq) \psi_n^*(\bq) }{E + \ui \eta - E_n}  
  = \frac{1}{\pi}  \lim_{\eta \to 0} \sum_n \frac{\eta}{(E-E_n)^2 + \eta^2}  =\sum_n \delta(E - E_n).
\end{align}
Here, the energy eigenfunctions were removed by equating the initial and final points and integrating over them, i.e., by taking a trace.

\subsubsection*{Periodic-orbit contributions}

To obtain an expression for the density of states in terms of trajectories,
 we apply (\ref{trace}) to the semiclassical Green function (\ref{Gscsimp}), and evaluate the integral over $\bq$,
 \begin{equation} \rho(E)\approx {\rm Re}\int \ud^2 q\sum_\gamma \frac{1}{\sqrt{ 2 \pi^3 \hbar^3 \dot{q}_{\parallel}^2  \, | M_{\gamma,12} | }} 
\exp \left( \frac{\ui}{\hbar} S_\gamma (\bq_b,\bq_a,E) - \ui \frac{\pi}{2} \xi_\gamma - \ui \frac{\pi}{4} \right) .
\end{equation} 
  using a stationary-phase approximation.
The stationary-phase condition
\begin{equation} 0 = \frac{\partial S_\gamma (\bq ,\bq,E)}{\partial \bq}=\left( \frac{\partial S_\gamma (\bq,\bq_0,E)}{\partial \bq}   +   \frac{\partial S_\gamma (\bq,\bq_0,E)}{\partial \bq_0}\right)_{\bq_0=\bq}
= \bp - \bp_0 
\end{equation}
means that the beginning and end of the trajectories coincide not only in position but also in momentum. Hence, the trajectories are {\bf periodic}. 

We now have to {\bf investigate whether the stationary points are isolated}, hence meeting the conditions for the stationary phase approximation.   They cannot be isolated along the parallel direction: For $\bq$ to satisfy the stationary-phase condition it must be on a periodic orbit, and moving along the direction of motion on that orbit leads to different points satisfying the same condition. However, for chaotic systems, the stationary points are isolated along the perpendicular direction. To explain this fact, it is helpful to consider the eigenvalues
 $m_1$ and $m_2$   of the stability matrix. These eigenvalues satisfy
\begin{equation}\label{Meval}
\det M = m_1 m_2 = 1, \qquad \Tr M = m_1 + m_2  ,\end{equation}
implying  \begin{equation}
m_{1,2} = \frac{1}{2} \left( \Tr M \pm \sqrt{(\Tr M)^2 - 4 } \right).  
\end{equation}
For chaotic systems, we have $|{\rm Tr}\, M|>2$,
leading to real mutually inverse eigenvalues with, say, $|m_1|>1$ and $|m_2|<1$. Initial deviations along the eigenvector associated with $m_1$ increase in modulus, signaling chaos. Moreover, a trajectory starting with a given perpendicular deviation from our periodic orbit cannot come back to the same phase-space point in the end, as this would require an eigenvalue 1. 

Hence, we can perform a {\bf one-dimensional stationary-phase approximation in $q_\perp$}. Using (\ref{derivs}), we can simplify the second derivative to give
\begin{equation}
\sigma=\frac{\partial^2 S_\gamma(\bq,\bq,E)}{\partial q_\perp^2}=\left(  \frac{\partial^2 S_\gamma(\bq,\bq_0,E)}{\partial q_{\perp}^2} + 2 \frac{\partial^2 S_\gamma(\bq,\bq_0,E)}{\partial q_{\perp} \partial q_{0\perp}} + \frac{\partial^2 S_\gamma(\bq,\bq_0,E)}{\partial q_{0\perp}^2} \right)_{\bq_0=\bq}=
\frac{M_{\gamma,22}-2+M_{\gamma,11}}{M_{\gamma,12}}=\frac{{\rm Tr}\, M_\gamma-2}{M_{\gamma,12}}.
\end{equation}  
The  stationary phase approximation for $q_\perp$ then yields 
\begin{align}
 \rho_{\rm osc} (E)
& = \Re \sum_p\int \! \frac{\ud q_\parallel}{\dot{q}_\parallel} \, \;  \frac{1}{\pi \hbar \, \sqrt{ | \Tr M_{p} - 2 | }} 
\exp \left( \frac{\ui}{\hbar} S_p(E) - \ui \mu_p \frac{\pi}{2} \right).
\end{align}
Here we have renamed our summation variables to $p$ as we now know that they correspond to periodic orbits. We have also introduced a subscript `osc' to indicate that in this case the stationary-phase approximation does not capture all relevant contributions but only the oscillatory ones - a fact that will be discussed further below. The phase depends on the {\bf Maslov index}
\begin{equation} 
\mu_p= \begin{cases} \xi_p , \qquad & \sigma > 0 , \\ \xi_p + 1 , \qquad & \sigma < 0 . \end{cases} 
\end{equation} 
The intuitive interpretation of the case distinction here is that $\xi_\gamma$ can change along a periodic orbit between two values that differ by one, depending on which point along the orbit is considered the start point.  The  Maslov index is the larger of these two values.

The {\bf integral over $q_{\parallel}$} can now be written as an integral over times along the orbit, noting that $\frac{dq_\parallel}{\dot q_\parallel}=dt$. As the integrand is constant, one would expect multiplication with the period $T_p$ of the periodic orbit. However, $p$ may involve multiple repetitions of a shorter `primitive' orbit. In this case, we only have to multiply with the period $T_p^{\rm prim}=T_p/r_p$ of the latter (where $r_p$ is the number of repetitions). This choice  can be justified because deformations of the orbit on any of the repetitions lead to the same mon-periodic trajectories. We set $r_p=1$ if $p$ is primitive. 
Altogether, we thus obtain 
\begin{align}
 \rho_{\rm osc} (E)
 = \frac{1}{\pi\hbar}\Re \sum_p  \frac{T_p^{\rm prim}}{\sqrt{ | \Tr M_{p} - 2 | }} 
\exp \left( \frac{\ui}{\hbar} S_p(E) - \ui \mu_p \frac{\pi}{2} \right),
\end{align}
which oscillates as the energy is varied.
Defining $F_p=\frac{\ue^{- \ui \mu_p \frac{\pi}{2}}}{\sqrt{ | \Tr M_{p} - 2 | }}$ we can abbreviate the result as
\begin{align}
 \rho_{\rm osc} (E)
 = \frac{1}{\pi\hbar}\Re \sum_p  F_p T_p^{\rm prim}\ue^{\frac{\ui}{\hbar} S_p(E)}.
\end{align}
In higher dimensions $| \Tr M_{\gamma} - 2 |$ has to
be replaced by $|\det (M - I)|$.

\subsubsection*{Weyl term}

The previous approximations do not correctly capture the contributions of direct trajectories from $\bq_0$ to $\bq$, as the corresponding action vanishes upon identifying the initial and final points, rendering the stationary-phase approximation meaningless. Instead, we use
that for $\bq_0$ close to $\bq$,
the direct trajectory   approaches that of a free particle with energy $E$
replaced by $E-V((\bq+\bq_0)/2)$. Its contribution to the Green function is then approximated by the corresponding modification of the free Green function
\begin{equation} 
G_{\rm direct}(\bq,\bq_0,E) = \frac{m}{2 \ui \hbar^2} \, H_0^{(1)} \left( \frac{\sqrt{2m(E-V)}}{\hbar} | \bq - \bq_0 | \right) \,  
\end{equation}
where $V=V((\bq+\bq_0)/2)$. The resulting contribution to the density of states is
\begin{equation}
\bar{\rho}(E)  \approx - \frac{1}{\pi} \lim_{\eta \to 0} \Im \int_{A(E)} \! \ud^2 q \: G_{\rm direct}(\bq,\bq,E+\ui\eta)
= \frac{m}{2 \pi \hbar^2}  \int_{A(E)} \! \ud^2 q = \frac{m A(E)}{2 \pi \hbar^2} . 
\end{equation}
Here $A(E)$ is the area in which $E > V(\bq)$ and we used  $J_0(0)=1$. This 
is a special case of the Weyl (or Thomas-Fermi) approximation, expressing the (locally averaged) 
density of states in terms of the volume of the energy shell,
\begin{equation} 
\bar{\rho}(E) = \frac{1}{(2 \pi \hbar)^f} \int \! \ud^f q \, \ud^f p \; \delta \left( E- H(\bq,\bp) \right) 
\end{equation}
where $H(\bq,\bp)$ is the classical Hamiltonian.

\subsubsection*{Final result}

Altogether, the  density of states is given by Gutzwiller's trace formula \cite{Gutzwiller-1970,Gutzwiller-1971}
\begin{align}
 \rho(E)\approx \bar\rho (E) +\rho_{\rm osc} (E)
 = \bar\rho (E) +\frac{1}{\pi\hbar}\Re \sum_p  F_p T_p^{\rm prim}\ue^{\frac{\ui}{\hbar} S_p(E)},
\end{align}
involving a local average followed by oscillations related to the classical periodic orbits.
This approximation holds in the limit where the classical (reduced) actions are much larger than $\hbar$ which is justified for {\bf large energies}. 

Gutzwiller's trace formula is not convergent as such but can be regarded as an asymptotic series or be made convergent by    considering other functions $\sum_n f(E_n)$ instead of $\rho(E)$. For compact Riemannian surfaces, 
a related rigorous trace formula was derived by Selberg \cite{Selberg-1956}.

In the following sections, we will use the trace formula to explain statistical features of quantum energy levels by relating them to fundamental statistical features of periodic orbits, following from classical chaos. This key application of the theory does not require any knowledge of the specific orbits. In addition, the trace formula is helpful to explain key features of the density of states by relating them to specific periodic orbits \cite{Gutzwiller-1973}. A final application of the trace formula is to resolve individual energy levels. This is challenging, as resolving neighboring energy levels requires orbits with periods up to the Heisenberg time $T_H=2 \pi \hbar \bar{\rho}(E)$, and the number of periodic orbits increases exponentially with their period. Methods to improve the numerical resolution of energy levels against these odds include cycle expansions \cite{Cvitanovic-1989} and harmonic inversion \cite{Main-1999}.

Berry and Tabor \cite{Berry-1976,Berry-1977} derived a corresponding result for  integrable systems, involving integration over tori of periodic orbits. For mixed chaotic-integrable systems one also has to deal with bifurcations of periodic orbits \cite{Schomerus-1997} and break-up of tori.

\subsubsection*{Example: Billiard systems}

It is instructive to specify the trace formula for chaotic billiard systems, i.e. potential-free areas surrounded by a boundary whose shape induces chaotic motion. If we  work in dimensionless units
$\hbar = 2 m = 1$, the energy is related to the wavenumber as $E = k^2$ and the semiclassical limit corresponds to $k \to \infty$.
For chaotoc billiards,
the trace formula 
then has the form
\begin{equation}
\rho(k) = \sum_n \delta(k - k_n) \approx \bar{\rho}(k) + \sum_p \frac{L_p^{\rm prim} \, \cos \left( k L_p - \ui \frac{\pi}{2} \mu_\gamma \right) }{\pi   \sqrt{|\Tr M_\gamma - 2|}} ,
\qquad \text{\color{black} where} \qquad \bar{\rho}(k) \approx \frac{A k}{2 \pi} .
\end{equation}
where $L_p$ is the length of the orbit $p$ and $L_p^{\rm prim}$ is its primitive length.
The stability matrices can be written as alternating products of two types of factors, corresponding to reflections and a straight-line path between two reflections.
Each reflection gives rise to a factor
\begin{equation}
\begin{pmatrix} -1 & 0 \\ \frac{2 k}{R \cos\beta} & -1 \end{pmatrix},\end{equation}
where $\beta$ is the angle of reflection and $R$ the radius of curvature (negative for concave boundaries). Each straight line of length $l$ leads to a factor  
\begin{equation}
 \begin{pmatrix} 1 & \frac{l}{k} \\ 0 & 1 \end{pmatrix}.
\end{equation}
When multiplying with an initial deviation, 
the rightmost factor in each product is applied first and hence should correspond to the first segment of the trajectory. 
The Maslov index has an additional contribution of twice the number of reflections at walls with Dirichlet boundary conditions.
For concave billiards, there are no conjugate points.

The trace formula for billiards was derived by Balian and Bloch \cite{Balian-1970,Balian-1972} but can also be recovered from Gutzwiller's trace formula.

\subsection{Spectral determinant}

\label{sec:det}

For some applications, it is preferable to work with the spectral determinant
\begin{equation}\label{det}
\Delta(E)=\det(E-\hat H)=\prod_n(E-E_n)	
\end{equation}
instead of the level density. Indeed, we will use this determinant to study spectral statistics in section \ref{sec:full}. Postponing a discussion of convergence for now, we can derive a semiclassical approximation of the spectral determinant using that
\begin{equation}\label{detint}
\Delta(E^+-\hat H)=\exp\left({\rm Tr}\ln(E^+-\hat H)\right)\propto
\exp\left({\rm Tr}\int^{E^+}\ud E'\frac{1}{E'-\hat H}\right)=\exp\left(\int^{E^+}\ud E'\int \ud^f q \,G(\bq,\bq,E')\right).
\end{equation}
We have already taken an integral of the Green function in the derivation of the trace formula. We thus obtain a very similar expression in the exponent here. Integration of the Weyl term gives an integrated level density $\bar N(E)$. 
If we expand the exponentiated orbit sum into a Taylor series, we obtain products of sums over orbits. These may be condensed into a sum over collections of orbits, also called {\bf pseudo-orbits}, which may include multiple repetitions of the same orbit.
For energies with a small positive imaginary part, we obtain  \cite{Berry-1990}
\begin{equation}
\label{det1}
\Delta(E^+)\propto
\ue^{-\ui\pi\overline{N}(E^+)}
\sum_{A} F_A(-1)^{n_A}\ue^{\ui S_A(E^+)/\hbar}.
\end{equation}
Here,  
$n_A$ indicates the number of orbits included in the pseudo-orbit $A$,
$S_A$ is the sum of actions,
and $F_A$ the products of the factors $F_p$.
If the energy is taken with a negative imaginary part, complex conjugation yields 
\begin{equation}
\Delta(E^-)\propto\ue^{\ui\pi\overline{N}(E^-)}
\sum_{A} F_A^*(-1)^{n_A}\ue^{-\ui S_A(E^-)/\hbar}.
\end{equation} 
Analogous reasoning for inverse spectral determinants leads to
\begin{align}\label{denominator}
\Delta(E^+)^{-1}&\propto
\ue^{\ui\pi\overline{N}(E^+)}
\sum_A F_A\ue^{\ui S_A(E^+)/\hbar}\notag\\
\Delta(E^-)^{-1}&\propto
\ue^{-\ui\pi\overline{N}(E^-)}
\sum_A F_A^*\ue^{-\ui S_A(E^-)/\hbar}
\end{align}

However, for non-inverted spectral determinants with real arguments, an improved  
approximation was derived by Berry and Keating by resummation  
\cite{Berry-1990,Keating-1992,Berry-1992}.
This approximation was motivated by the Riemann-Siegel formula for the Riemann zeta function and is hence referred to as the {\bf Riemann-Siegel lookalike}. In this formula, the contribution of pseudo-orbits with cumulative periods larger than half of the Heisenberg time is replaced by the complex conjugate of the contribution from shorter pseudo-orbits,
\begin{equation}
\label{rs}
\Delta(E)\propto
\ue^{-\ui\pi\overline{N}(E)}
\sum_{A\;(T_A<T_H/2)} F_A(-1)^{n_A}\ue^{\ui S_A(E)/\hbar}+{\rm c.c.}
\end{equation}
 In a more refined version, the sum is smoothly truncated using a complementary error function \cite{Berry-1992}.
The Riemann-Siegel lookalike explicitly incorporates the fact that the energy levels are real, and the same applies to the spectral determinant for real arguments. A heuristic derivation can be given by postulating that (\ref{det1}) is equal to its complex conjugate for real energies and taking a Fourier transform with a restricted range on both sides. The resulting identity leads to the desired identification of contributions from $T_A>T_H/2$ with complex-conjugated contributions from $T_A<T_H/2$ on the two sides, where
 the threshold arises from the derivative of the phase, $-\pi\bar\rho+T_A/\hbar=(-T_H/2+T_A)/\hbar$.

The spectral determinant as defined in (\ref{det}) typically diverges. This issue can be avoided if one works with a {\bf regularised} determinant of the form
\begin{equation}
\Delta_{\rm reg}(E) = \prod_{n=1}^\infty A(E,E_n) \, (E - E_n).
\end{equation}
Berry and Keating stated their result in a regularized setting, involving a factor $B(E)$ that absorbs the proportionality factor that we left unspecified here, as well as the effects of regularization. 
For billiards, a suitable choice is  \cite{Sieber-2007}
\begin{equation}
	A(E,E_n)=
- \frac{1}{E_n} \ue^{E/E_n}.
\end{equation}

\section{Foundations of spectral statistics}
  
\subsection{Correlation function and spectral form factor}

We are now ready to apply periodic-orbit theory to study the statistical behavior of quantum energy levels. The spectra of fully chaotic quantum systems have universal statistical   properties in line with phenomenological predictions from random matrix theory (RMT). In particular, the energy levels show a tendency to repel. Based on numerical evidence, this universality was conjectured by Bohigas, Giannoni, and Schmit \cite{Bohigas-1984}, and related work was reported in \cite{McDonald-1979, Casati-1980}.

 Key quantities to describe spectral statistics \cite{Haake-2018} include the two-point correlation function of    (the oscillatory part of) the level density 
 \begin{equation}
R(\eps)=\frac{1}{\bar\rho^2}\left\langle\rho_{\rm osc}\left(E+\frac{\eps}{2\pi\bar\rho}\right)
\rho_{\rm osc}\left(E-\frac{\eps}{2\pi\bar\rho}\right)\right\rangle
\end{equation} 
as well as its Fourier transform, the spectral form factor
\begin{equation}
K(\tau)=\frac{1}{\pi}\int_{-\infty}^\infty  R(\eps) \,\ue^{2\ui\eps\tau}\ud\eps.
\end{equation} 
Here, $\eps$ plays the role of a dimensionless energy difference and $\tau$ is a dimensionless time variable conjugate to $\eps$. The average is taken over the center energy $E$ as well as a small window of $\eps$ to obtain a smooth quantity.  
To motivate $R(\eps)$, we note that it is directly related to the correlation function of the full level density as
\begin{equation}
R(\eps)+1=\frac{1}{\bar\rho^2}\left\langle\rho \left(E+\frac{\eps}{2\pi\bar\rho}\right)
\rho \left(E-\frac{\eps}{2\pi\bar\rho}\right)\right\rangle.
\end{equation} 
This relation can be checked if we use
 that products of one Weyl term and one oscillatory level density average to zero.
In case of a constant level density, we can then write $R(\eps)+1$ in terms of the energy levels $E_j$,
\begin{equation}
R(\eps)+1=\left\langle\sum_{j,k}\delta\left(\eps-\pi\bar\rho(E_j-E_k)\right)
\right\rangle.
\end{equation}
It hence gives the likelihood that $\eps$ arises as a (scaled) difference between energy levels. Other  quantities describing spectral statistics include higher-order correlation functions as well as the distribution of nearest neighbor level spacings.

\subsection{Random matrix theory}

\begin{figure}[h]
\centering
\includegraphics[width=1\textwidth]{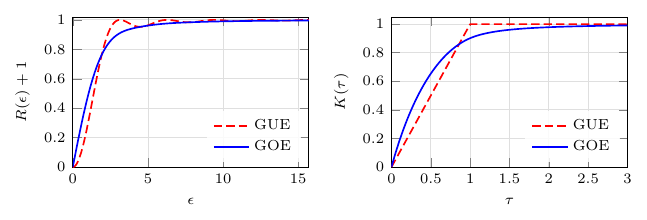} 
\caption{Correlation function and spectral form factor for the GUE and GOE}
\label{plots}
\end{figure}

 The random matrix predictions are obtained by modeling the Hamiltonians of chaotic systems as large matrices and averaging over all matrices compatible with the general features of the Hamiltonian. In this volume, a detailed review is given in \cite{Kieburg-2026}. In the Gaussian random-matrix ensembles the average is taken with a   weight proportional to $\exp\left(-\alpha\,{\rm tr}H^2\right)$ where $\alpha$ is constant. Assuming that our Hamiltonian has no symmetries except possibly time reversal invariance (TRI), we obtain the following 
  ensembles: \begin{itemize} 
\item  In the absence of even TRI, all we know about the Hamiltonian is that it is Hermitian.  We thus average over Hermitian matrices with a Gaussian weight. This ensemble is referred to as the {\bf Gaussian Unitary Ensemble (GUE)}, due to the invariance of the Gaussian weight under unitary transformations. 
\item Many systems are invariant under time reversal. Conventional time reversal symmetry, on which we focus here, implies Hamiltonians even in the momentum. The time-reversed of a classical trajectory (with flipped sign of the momentum as well as time) also satisfies the classical equations of motion. The corresponding quantum Hamiltonian is real and commutes with a complex conjugation operator, which can be regarded as a time reversal operator $T$ and squares 1. The corresponding ensemble is the {\bf Gaussian Orthogonal Ensemble (GOE)} of real symmetric matrices with a Gaussian weight. The same ensemble arises if the Hamiltonian commutes with a different antiunitary operator that squares to $1$.
\item  It is also possible to have TRI with a time-reversal operator squaring to $-1$. This typically arises for spin systems with half-integer spin. One can show that the appropriate matrices can be written in quaternion-real form, involving $2\times 2$ blocks of the type $\left(\begin{matrix}z&w\\-w^*&z^*\end{matrix}\right)$. The Gaussian weight is now left invariant by symplectic transformations, and we refer to this ensemble as the {\bf Gaussian Symplectic Ensemble (GSE)}. 
\end{itemize}
If we also allow for antiunitary and/or unitary operators anticommuting with the Hamiltonian (expressing charge conjugation and chiral \cite{Verbaarschot-1994} symmetries), this threefold classification of ensembles can be extended into Altland's and Zirnbauer's 'tenfold way' \cite{Altland-1997}. 
  Our presentation will focus on the GUE and the GOE. For these, the random-matrix predictions are 
\begin{align}
R_{\mathrm{GUE}}(\eps)
&=
-\left(\frac{\sin\eps}{\eps}\right)^2
= {\rm Re}\left[
-\frac{1}{2\eps^2}
+\frac{\ue^{2\ui\eps}}{2\eps^2}\right]\\
K_{\mathrm{GUE}}(\tau)
&= 
\begin{cases}
\tau, & 0 \le \tau \le 1\\[3pt]
1, & \tau \ge 1
\end{cases} \end{align}
and \begin{align}
R_{\mathrm{GOE}}(\eps)
&=
-\left(\frac{\sin\eps}{\eps}\right)^2
-
\left(\frac{d}{d\eps}\frac{\sin\eps}{\eps}\right)
\left(\frac{\pi}{2}-\operatorname{Si}(\eps)\right) \notag\\
&={\rm Re}\left[-\frac{1}{\eps^2}+\sum_{n=3}^\infty\frac{(n-3)!(n-1)}{2(\ui\eps)^{n}}+
\sum_{n=4}^\infty\frac{(n-3)!(n-3)}{2(\ui\eps)^{n}}\ue^{2\ui\eps}\right]\\
K_{\mathrm{GOE}}(\tau)
&=
\begin{cases}
2\tau-\tau\ln(1+2\tau), & 0 \le \tau \le 1\\[3pt]
2-\tau\ln\!\left(\dfrac{2\tau+1}{2\tau-1}\right), & \tau \ge 1.
\end{cases}
\end{align} The correlation functions are split into non-oscillatory contributions as well as oscillatory contributions involving factors $\ue^{2\ui\eps}$.
In case of the GOE, both can be organised as asymptotic series in $\frac{1}{\eps^n}$.
Upon Fourier transformation, the non-oscillatory contributions give rise to an expansion of the form factor in powers of $\tau$. 
We note that such contributions also arise for odd $n$, even though the corresponding terms vanish in $R(\eps)$ for real $\eps$. The technical justification for this is that $R(\eps)$ can be viewed as the real part of a complex correlator (also with complex $\eps$), which has the advantage of being recoverable from its series expansion by Borel summation \cite{Haake-2018}.
The leading terms in the  resulting expansion of $K(\tau)$ are $2\tau-2\tau^2+2\tau^3-\frac{8}{3}\tau^4$. Conversely, Fourier transformation of the oscillatory terms gives contributions proportional to $(\tau-1)^{n-1}\Theta(\tau-1)$, which lead to the distinct formulas for $\tau\geq 1$. 

The resulting correlation functions and form factors are plotted in Fig. \ref{plots}. We see that $R(\eps)+1$, which gives the likelihood for $\eps$ to arise as a scaled energy difference, vanishes for $\eps\to0$. Hence, small energy differences are suppressed, as expected.
   
 We proceed to summarize the semiclassical literature, showing that individual chaotic systems are faithful to these random-matrix predictions. We will use the Gutzwiller trace formula and the Riemann-Siegel lookalike formula, as well as classical properties of the chaotic dynamics.    
   
\subsection{ Classical chaos }

\begin{figure}[h]
\centering
\includegraphics[width=.9\textwidth]{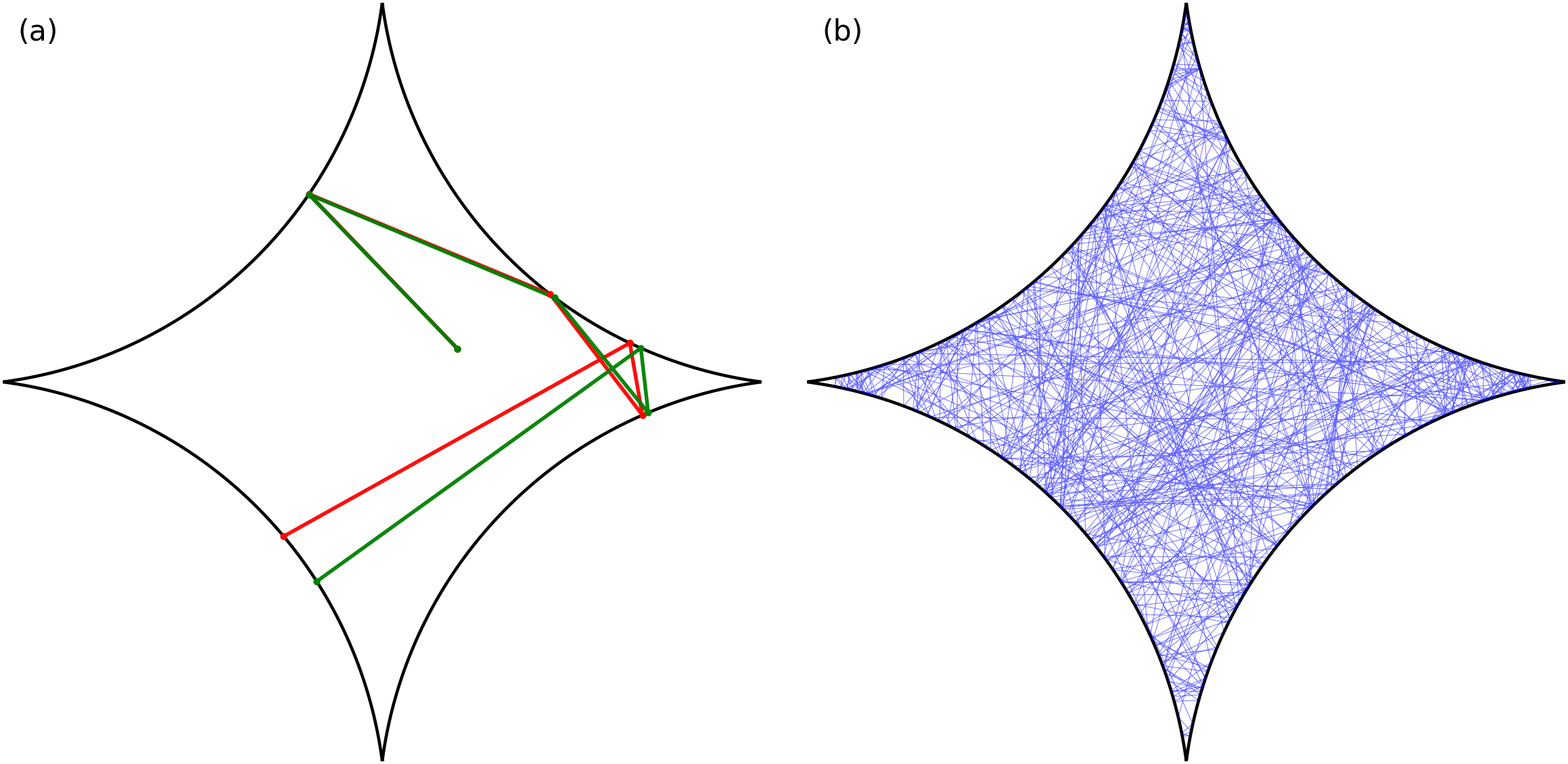}
\caption{(a) Separation of two trajectories in the diamond billiard, illustrating sensitive dependence on initial conditions. (b) Long trajectory in the diamond billiard (50 reflections), illustrating ergodicity.}
\label{classical}
\end{figure}

We give a short overview of classical chaos, assuming for simplicity a two-dimensional position space. For a detailed review, we refer to \cite{Tomsovic-2026}.
 
Fully chaotic systems display a {\bf sensitive dependence on initial conditions} (or more formally, {\bf hyperbolicity}). To define this property, we consider trajectories starting at closeby points on the same energy shell. The separation between these points can then be decomposed into an unstable component $u$, a stable component $s$ and a time difference $\Delta t$. It will be advantageous for the following to normalize the local basis vectors associated to the stable and unstable directions such that their symplectic product satisfies $\boldsymbol{e}_u\wedge\boldsymbol{e}_s=1$. If the starting points are taken in a suitable Poincaré surface of section (e.g. perpendicular to the trajectories in position space), the time difference $\Delta t$ can be taken as zero. 
 If we now follow the relative motion of the trajectories and move the surface of section along, the unstable separation increases exponentially in time, whereas the stable separation decreases exponentially. The long-time behavior of these separations is given by
\begin{equation}\label{lyapunov}
u(t)\approx u(0)\ue^{\lambda t},\quad\quad s(t)\approx s(0)\ue^{-\lambda t}
\end{equation}
where $\lambda$ is called the Lyapunov exponent.
(For finite trajectories, we described analogous behaviour when we considered the eigenvalues of the stability matrices in (\ref{Meval}).)
 If the initial separation is a combination of a stable and a non-zero unstable components, the growth of the unstable component eventually dominates, and the trajectories separate from each other. An example is shown in Fig. \ref{classical}(a).

Another key classical property of chaotic systems is {\bf ergodicity}. In an ergodic system, long classical trajectories uniformly explore the corresponding energy shell. For example, in the case of a billiard, the interior is filled uniformly and there are no preferred directions, as illustrated in Fig. \ref{classical}(b).
More formally, in the long term limit, the average of a phase space observable over a long trajectory converges to the energy shell average except for exceptional trajectories, which are altogether of measure zero. For hyperbolic systems, the uniform measure on the energy shell employed here can locally be written as  
$\ud s\, \ud u\, \ud\Delta t$.

For the following work, we need a condition that is related to but stronger than ergodicity. To state this condition, we consider the Frobenius-Perron operator $P(t)$, which describes the classical evolution of phase-space densities over time $t$, with each phase space point moving according to Hamilton's equations. $P(t)$ has an eigenvalue decomposition with eigenvalues that decay exponentially over time, as in $\ue^{-\gamma_j t}$, where $\gamma_j$ are known as Ruelle-Pollicot resonances. For an ergodic system, one of these resonances is zero, and the corresponding eigenfunction of the Frobenius-Perron operator is given by the uniform density on the energy shell. Now we require a {\bf spectral gap}, i.e. that all other resonances have real parts bounded away from zero.

For systems with a spectral gap, sums over long periodic orbits can be evaluated using the {\bf Hannay-Ozorio de Almeida sum rule} \cite{Hannay-1984}
\begin{equation}
\sum_p |F_p|^2 \left( T_p^{\rm prim}\right)^2\delta(T_p-T)\approx T.
\end{equation}
Here, the approximation holds for large $T$. The summand is the square of the orbit weight arising in the Gutzwiller trace formula. The rule involves an average over a small window of $T$ left implicit in our notation. (The rule can be derived by expressing the trace of the Frobenius-Perron operator as a sum over periodic orbits and noting that only the vanishing resonance contributes for large times.) For our present purposes, we can use the primitive period and the period interchangeably, since orbits involving repetitions of shorter orbits are exponentially suppressed compared to the others. For our applications, it is furthermore useful to rewrite the sum rule in terms of an integral over a function $g$ depending on the period, 
\begin{equation}\label{sumrule}
\sum_p |F_p|^2 g(T_p)\approx\int_0^\infty \frac{\ud T}{T}g(T).
\end{equation}

\subsection{Double sum over orbits}

We are now equipped to evaluate the correlation function $R(\eps)$ using semiclassical methods. We will start with the leading orders in $\frac{1}{\eps}$. To simplify a later extension to the full correlation function, we will express  $R(\eps)$ in terms of the autocorrelation function of the integrated level density $N(E)$,
\begin{equation}
A(\eps_1,\eps_2)=\left\langle N_{\rm osc}\left(E+\frac{\eps_1}{2\pi\bar\rho}\right)N_{\rm osc}\left(E-\frac{\eps_2}{2\pi\bar\rho}\right)\right\rangle
\end{equation}
as
\begin{equation}\label{RfromA}
R(\eps)=-4\pi^2\frac{\partial^2}{\partial\eps_1\partial\eps_2}A(\eps_1,\eps_2)\big|_{\eps_1=\eps_2=\eps}.
\end{equation}
 We can write $N_{\rm osc}(E)$ as 
\begin{equation}
N_{\rm osc}(E)=\frac{1}{\pi}{\rm Im}\sum_p F_p(E) \exp\left(\frac{\ui}{\hbar}S_p(E)\right)
\end{equation}
where we use that $\frac{d S_p}{dE}=T_p$, see (\ref{Sderiv}), which agrees with $T^{\rm prim}_p$ except for an exponentially suppressed number of orbits. We then obtain the double sums over periodic orbits 
\begin{align}\label{Adoublesum}
A(\eps_1,\eps_2)=&\frac{1}{2\pi^2}{\rm Re}\left\langle\sum_{p,p'} F_p F_{p'}^*\exp\left(\frac{\ui}{\hbar}\left(S_p-S_{p'}\right)\right)\exp\left(\frac{\ui(T_p\eps_1+T_{p'}\eps_2)}{T_H}\right) \right\rangle\notag\\
-&\frac{1}{2\pi^2}{\rm Re}\left\langle\sum_{p,p'} F_p F_{p'}\exp\left(\frac{\ui}{\hbar}\left(S_p+S_{p'}\right)\right)\exp\left(\frac{\ui(T_p\eps_1-T_{p'}\eps_2)}{T_H}\right)\right\rangle
\end{align}
where we Taylor expanded the action to first order,  neglected the effect of the energy increments on $F_p$ and $F_{p'}$, dropped the energy arguments $E$, and introduced the {\bf Heisenberg time} $T_H=2\pi\hbar\bar\rho$. Crucially, our expression involves an average over the center energy $E$. As $E$ varies, the actions vary, and the exponents oscillate rapidly. Hence, most of the contributions to $A(\eps_1,\eps_2)$ are expected to approximately average to zero. This is avoided only if two actions nearly compensate, i.e., in the first summand for $S_{p'}\approx S_p$. 
For this situation, we can drop the second summand and approximate the first summand assuming that $F_{p'}$ and $T_{p'}$ are (near) identical to the properties of $p$. We can also simplify the derivatives leading to $R(\eps)$ as the contributions now depend only on $\eps_1+\eps_2$. This gives  
\begin{equation}\label{doublesum}
R(\eps)\approx\frac{1}{2}\frac{\partial^2}{\partial\eps^2} {\rm Re}\left\langle\sum_{p,p'} |F_p|^2 \exp\left(\frac{\ui}{\hbar}\left(S_p-S_{p'}\right)\right)\exp\left(\frac{2\ui T_p\eps}{T_H}\right)  \right\rangle.
\end{equation}

\subsection{Diagonal approximation}

An important contribution to the double sum arises if 
the actions are identical, due to the orbits being identical or mutually time reversed. This is the idea of the diagonal approximation (see Berry \cite{Berry-1985}, Hannay and Ozorio de Almeida \cite{Hannay-1984}). The resulting single sum over orbits \begin{equation}
R_{\rm diag}(\eps)=
-\frac{\kappa}{2}\frac{\partial^2}{\partial\eps^2} \left\langle\sum_p |F_p|^2 \exp\left(\frac{2\ui T_p\eps}{T_H}\right)\right\rangle,
\end{equation}
with $\kappa=1$ for systems without time reversal invariance (TRI) and $\kappa=2$ for systems with TRI,
can then be evaluated using the sum rule (\ref{sumrule}) to give 
\begin{equation}
R_{\rm diag}(\eps)=-\frac{\kappa}{2}\frac{\partial^2}{\partial\eps^2} \int_0^\infty \frac{\ud T}{T}\exp\left(\frac{2\ui T\eps}{T_H}\right)=-\frac{\kappa}{2\eps^2}.
\end{equation}
This clarifies the semiclassical origin of leading terms in the random-matrix prediction, for the GUE and the GOE. For the GOE we expect a series expansion with infinitely many further contributions. For the GUE we just need one additional oscillatory term, as well as confirmation that potential further contributions vanish or compensate.

\begin{BoxTypeA}

\noindent 
{\bf Exercise:} Which result do we expect for systems that also have a geometrical symmetry, such as  reflection symmetry or symmetry w.r.t. rotations by $\frac{2\pi}{n}$?

\end{BoxTypeA}

\begin{BoxTypeA}

\noindent 
{\bf Exercise:} Let $D_n$ be the diagonal approximation to $|K(\bq,\bq,t)|^{2n}$, $=1,2,\dots$, capturing contributions that do not oscillate rapidly. Express all $D_n$ in terms of $D_1$, for systems with and without TRI. 
\end{BoxTypeA}

\subsection{Pairs of orbits differing in a 2-encounter} 

\label{sec:sr}

 Additional contributions to the double sum are expected to arise from pairs of orbits whose actions are not identical (modulo time reversal) but similar, and hence still allow for constructive interference. 
 The study of such correlations between the actions of periodic orbits was initiated in \cite{Argaman-1993}.
 Although the spectrum of periodic orbit actions is to a large degree Poissonian \cite{Cohen-1998,Smilansky-2003}, signaling statistical independence, one can nevertheless show that the spectral statistic according to the predictions of RMT requires nontrivial correlations, and an associated classical action correlation function was derived from RMT in \cite{Argaman-1993}.
 
\begin{figure}[h]
\centering
\includegraphics[width=.7\textwidth]{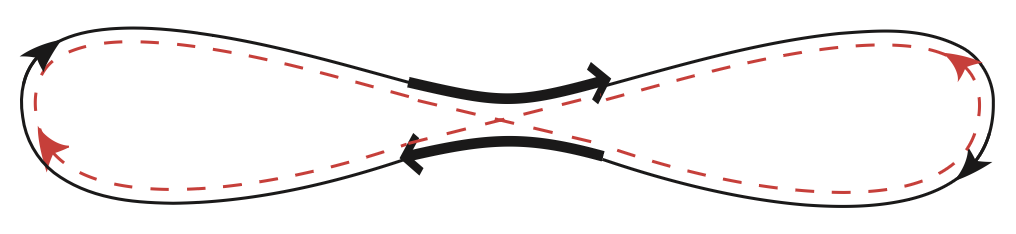}
\caption{Orbits differing in a 2-encounter: Schematic sketch.}
\label{sr}
\end{figure}

\begin{figure}[h]
\centering
\includegraphics[width=.8\textwidth]{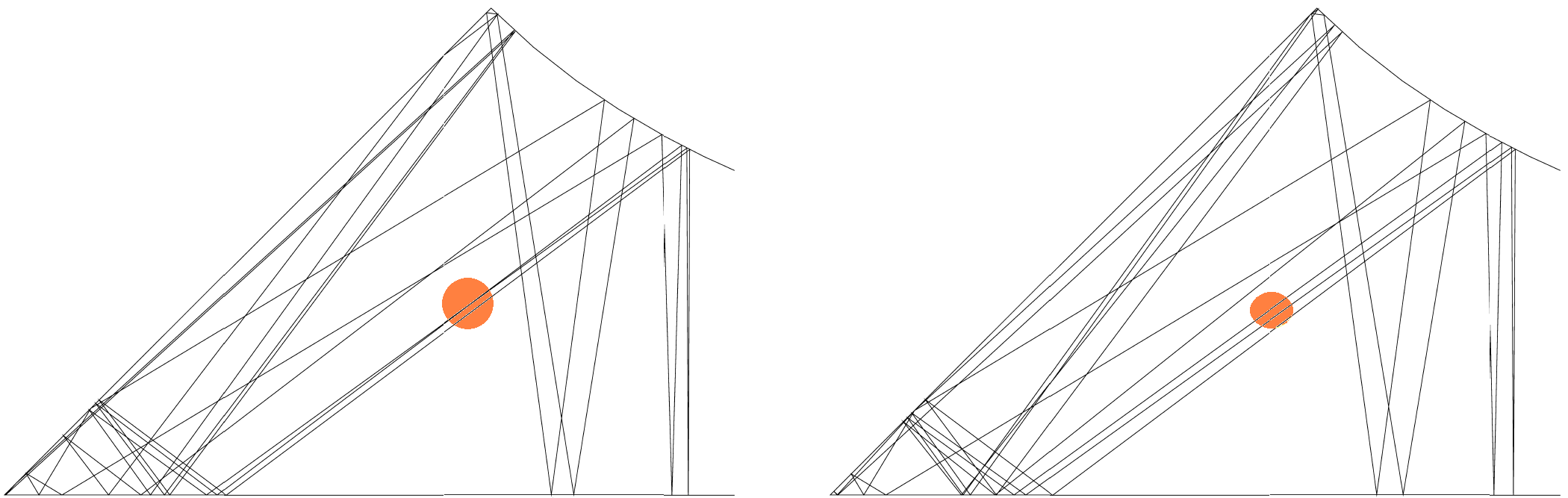}
\caption{Orbits differing in a 2-encounter: Example in a hyperbola billiard. The right part of the billiard is cut off. On the left, a crossing is highlighted in orange. A third piece of trajectory happens to traverse the orange circle is well, but is not relevant for our considerations. On the right, the crossing is replaced by an avoided crossing.}
\label{srbilliard}
\end{figure}

Indeed, a systematic 
 correlation mechanism was identified in \cite{Sieber-2001, Sieber-2002} explaining the next to leading orders of $R(\eps)$ and $K(\tau)$ for systems with TRI. It involves pairs of orbits $p$ and $p'$ where one orbit contains a self-crossing in a position space with a small angle and its partner narrowly avoids this crossing. In $p'$, a part of $p$ is almost reverted in time, suggesting that these orbit pairs can exist only for time-reversal invariant systems. A schematic of such an orbit pair is shown in Fig. \ref{sr}, and an example in a billiard is shown in Fig. \ref{srbilliard}.

It is helpful to describe the formation of these orbit pairs in phase-space language. We then say that $p$ involves an {\bf encounter} (more precisely, a 2-encounter) where two {\bf stretches} of the orbit are almost mutually time-reversed. The partner orbit $p'$ also involves such an encounter, but the endpoints of the stretches are connected in an approximately opposite way. The parts of the orbits outside the encounters are referred to as {\bf links}. The orbits approximately coincide in one link, whereas the other link is approximately time reversed.

\begin{figure}[h]
\centering
\includegraphics[width=.9\textwidth]{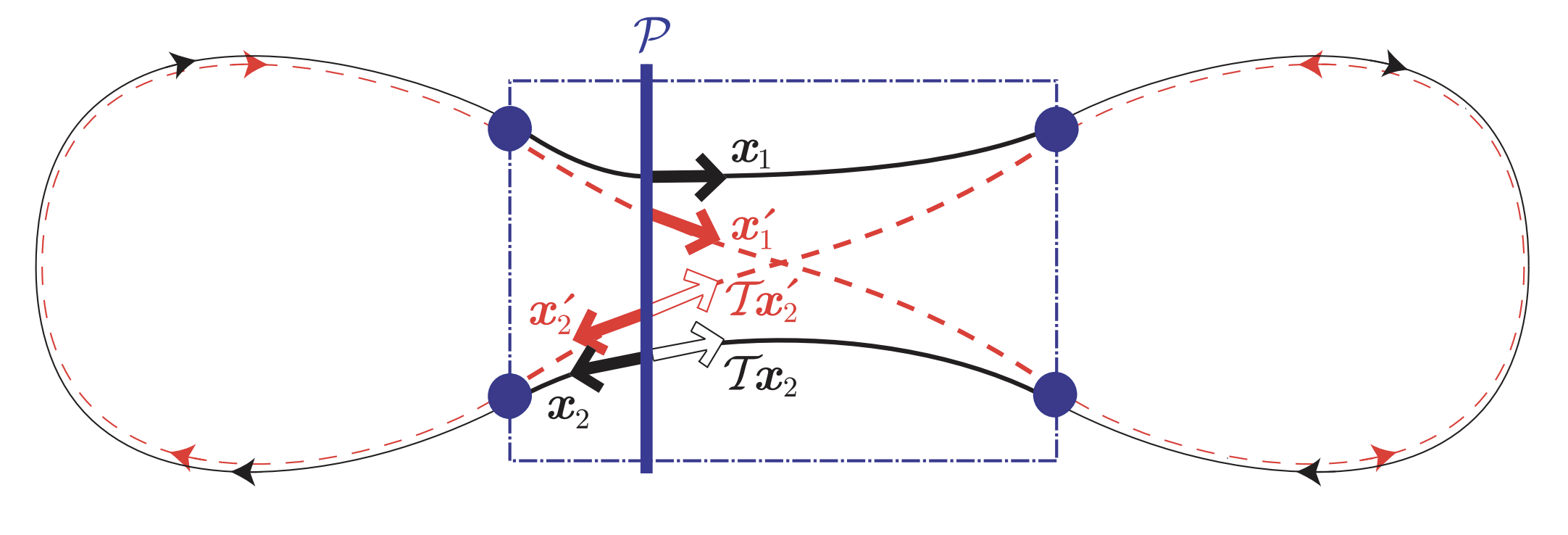}
\caption{Orbits differing in a 2-encounter: Illustration of the Poincar{\'e} section and phase-space points used in the text.}
\label{srconstruction}
\end{figure}

If we place a Poincar\'e surface of section inside the encounter. The orbit $p$ pierces through this section in the phase space points $\vx_1$ and $\vx_2$ (indicated by the positions and arrow directions in Fig. \ref{srconstruction}) which are close to being time reversed. Following \cite{Spehner-2003, Turek-2003}, the closeness of the encounter can then be characterized by splitting the deviation between the time reversed of  $\vx_2$ (indicated by $T\vx_2$) and $\vx_1$ into {\bf stable and unstable components}
\begin{equation}
T\vx_2-\vx_1=s\,\ve_s+u\,\ve_u
\end{equation}
where the basis vectors can be taken at $\vx_1$, for example. (Note that we are assuming a two-dimensional system; for higher-dimensional systems, one has to introduce several stable and unstable components \cite{Turek-2005}.)
The corresponding piercing points of $p'$, denoted by $\vx_1'$ and $\vx_2'$  can be determined using the properties of the stable and unstable directions. For example, the deviation of $T\vx_2$ of $\vx_1'$ from  must be approximately along the stable direction, as the orbit parts following in Fig. \ref{srconstruction} approach. In contrast, the deviation of $\vx_1'$ from $\vx_1$ must be along the unstable direction, as the orbits approach when going backward from these points.

One can show that the {\bf difference between the actions} of $p$  and $p'$ is in the leading order given by \cite{Spehner-2003, Turek-2003}
\begin{equation}
\Delta S = S_p-S_{p'}=su.
\end{equation} 
This is reassuring because if we shift the Poincar\'e section through the encounter, $s$ decreases and $u$ increases according to (\ref{lyapunov}), but the action difference remains the same. 
Moreover, if we formally define the encounter as the region where $|s|$ and $|u|$ remain below an arbitrary small constant c, its {\bf duration} can be written as \begin{equation}
t_{\rm enc} = \frac{1}{\lambda}\ln\frac{c}{|s|}+\frac{1}{\lambda}\ln\frac{c}{|u|}=\frac{1}{\lambda}\ln\frac{c^2}{|\Delta S|}.
\end{equation} 
   
\subsubsection*{Statistics of encounters}   
To determine the contribution of orbit pairs as in figure to $R(\eps)$, we have to determine the {\bf number $P_T(\Delta S)\ud\Delta S$ of encounters} in a long orbit of period $T$, such that the change of connection inside the encounter leads to a  partner orbit with action difference $\Delta S$. We proceed to informally sketch the evaluation of this quantity.
As a first step,
let us choose a fixed point on the orbit $p$ as $\vx_1$, and consider a perpendicular Poincar\'e section at this point. 
The formation of an encounter requires a second, almost mutually time-reversed piercing through this section.
One can show that the probability for such a piercing to occur in a time interval $\ud t$ with stable and unstable separations in intervals $\ud s$ and $\ud u$ is 
\begin{equation}
\frac{\ud s\,\ud u\,\ud t}{\Omega},
\end{equation}
where $\Omega$ is the volume of the energy shell. (This result holds due to what we said in section 2.3 about the measure on the energy shell.) 
We now have to integrate over $t$ to consider all possible piercings through our Poncar\'e section. In addition, a further integral over the orbit has to be taken to account for all possible placements of $\vx_1$. One of these integrals leads to multiplication with $T$. When evaluating the second integral, we have to make sure that the two piercing points do not come too close. This is because the orbit must keep space for two links with non-vanishing durations $t_1$ and $t_2$, as well as two non-overlapping encounter stretches, each of duration $t_{\rm enc}$. This is achieved by multiplying with   
\begin{equation}\label{linkint}
\int \ud t_1 \int \ud t_2\,\delta\left(t_1+t_2+2t_{\rm enc}-T\right),
\end{equation}
i.e. we integrate over the two link durations subject to the constraint $t_1+t_2+2t_{\rm enc}=T$. The integral is easily evaluated as $T-2t_{\rm enc}$, however it is advantageous for later generalization to avoid the step.   
   
Importantly, the same
partner orbit arises wherever we place the Poincar\'e section inside the encounter of duration $t_{\rm enc}$. Also, either of the two stretches of the encounter may be chosen as the first stretch containing $\vx_1$. To avoid overcounting, we therefore have to divide by $2t_{\rm enc}$. This leads to 
\begin{equation}\label{w}
w_T(s,u)=\frac{T\int \ud t_1 \int \ud t_2\,\delta\left(t_1+t_2+2t_{\rm enc}-T\right)}{2\Omega t_{\rm enc}}
\end{equation}
as a suitable weight associated to stable and unstable components $s$ and $u$. As the encounter is properly characterized by the action difference $\Delta S=su$ instead of the factors $s$ and $u$ we then write 
\begin{equation}
P_T(\Delta S)=\int \ud s\int \ud u\, w_T(s,u)\,\delta(\Delta S-su).
\end{equation}  

\subsubsection*{Contribution to the correlation function}

\label{sec:srcontrib}
 
We are now prepared to determine the contribution of the correlation function arising from orbit pairs differing in a 2-encounter.
We can evaluate the sum over $p$ in (\ref{doublesum}) using the sum rule (using that $F_{p'}\approx F_p$), and express the sum over $p'$  differing from $p$ in encounters using $P_T(\Delta S)$. As changing connections inside the encounter leads to two different mutually time-reversed partner orbits, we also have to multiply by 2. Altogether,
this gives
\begin{equation}
R_{\rm 2enc}(\eps)=-\frac{1}{2}\frac{\partial^2}{\partial\eps^2}{\rm Re}\int_0^\infty \frac{\ud T}{T}2
\int \ud\Delta S  P_T(\Delta S)\exp\left(\frac{\ui}{\hbar}\Delta S\right)\exp\left(\frac{2\ui T\eps}{T_H}\right).
\end{equation}
If we  cancel the $T$'s in (\ref{sumrule}) and (\ref{w}) and evaluate the integrals over $T$ and $\Delta S$ first, we obtain.
\begin{equation}
R_{\rm 2enc}(\eps)=-\frac{1}{2}\frac{\partial^2}{\partial\eps^2}{\rm Re}\int \ud t_1\int \ud t_2\int \ud s\int \ud u\frac{1}{\Omega t_{\rm enc}}	\exp\left(\frac{\ui}{\hbar}su\right)\exp\left(\frac{2\ui (t_1+t_2+2t_{\rm enc})\eps}{T_H}\right)
\end{equation}
This expression factorizes nicely into contributions associated with the links and the encounter. 
The factors can be written in a particularly compact form if we combine them with mutually canceling Heisenberg times $T_H$. 
The {\bf links} $j=1,2$ then yield factors 
\begin{equation}
\frac{1}{T_H}\int_0^\infty \ud t_j	\exp\left(\frac{2\ui t_j\eps}{T_H}\right)=\frac{1}{2\ui\eps},
\end{equation}
where the contribution of the upper integration limit vanishes if we take the energy increments with a small positive imaginary part.
The {\bf encounter} gives
\begin{equation}
	I=T_H^2\int \ud s\int \ud u\frac{1}{\Omega t_{\rm enc}}	\exp\left(\frac{\ui}{\hbar}su\right)\exp\left(\frac{4\ui t_{\rm enc}\eps}{T_H}\right).
\end{equation}
It is now helpful to Taylor expand the final exponential in powers of $t_{\rm enc}$. As shown in \cite{Muller-2009}, only the linear term in this expansion is relevant. 
For the constant term, the integral vanishes, and all higher-order terms are negligible in the semiclassical limit.
The integral of the relevant term can be easily evaluated as the two occurrences of $t_{\rm enc}$ cancel. We thus obtain
\begin{equation}
I=\frac{T_H}{\Omega}\underbrace{\int \ud s\int \ud u \exp\left(\frac{\ui}{\hbar}su\right)}_{=2\pi\hbar}4\ui\eps =4\ui\eps 
\end{equation}
where we used that $T_H=2\pi\hbar\bar\rho$ and $\bar\rho=\frac{\Omega}{(2\pi\hbar)^2}$.
The final result is therefore
\begin{equation}
R_{\rm 2enc}(\eps)=-\frac{1}{2}\frac{\partial^2}{\partial\eps^2}{\rm Re}	\frac{4\ui\eps 	}{(-2\ui\eps)^2}=-{\rm Re}\frac{\ui}{\eps^3},
\end{equation}
in agreement with the second term in the random-matrix prediction for time-reversal invariant systems. Even though this result vanishes for real $\eps$, it is crucial as upon Fourier transform it leads to a contribution $-2\tau^2$ to the spectral form factor, as required for agreement with the GOE.

Our present derivation is similar to \cite{Muller-2011} and optimized for extension in section \ref{sec:full}. We refer to \cite{Sieber-2001,Sieber-2002} for the introduction of these orbit pairs in systems of constant negative curvature and to \cite{Spehner-2003,Turek-2003,Muller-2003,Turek-2005} for generalizations.

We stress that the result arises from a sub-leading term in $I$ as the leading term in $I$ gives zero after integration. This sub-leading term corresponds to the correction $-2t_{\rm enc}$ in the expression $T-2t_{\rm enc}$ in (\ref{linkint}) as well as a logarithmic correction to the crossing distribution in \cite{Sieber-2001}.
It indicates a suppression of the systematic correlation mechanism for particularly small action differences.
 Ultimately, it arises as the two encounter stretches must be separated by a non-vanishing link. For some systems, such as billiards on surfaces of negative curvature \cite{Sieber-2001}, encounters failing this condition cannot exist. For others, they can exist due to an almost self-retracing reflection, but it is not possible to construct a partner orbit \cite{Muller-2003}. 

\section{Full correlation function}

\label{sec:full}

 To resolve higher orders $\frac{1}{\eps}$, one has to consider more complex orbit pairs based on the same general principle \cite{Heusler-2004}. Using these orbit pairs, the full expansion in $\frac{1}{\eps}$ was recovered in \cite{Muller-2004,Muller-2005}. However, the oscillatory terms and hence the behavior of the form factor for $\tau>1$ could only be obtained in a changed framework \cite{Heusler-2007,Keating-2007,Muller-2009,Haake-2018}  based on a generating function and the Riemann-Siegel lookalike formula introduced in subsection \ref{sec:det}. We proceed to give an overview of this framework, which also brings in sets of larger numbers of orbits rather than only pairs. It is tempting to interpret this approach as a refinement of the semiclassical approximation incorporating additional quantum-mechanical information. It is currently not known whether this refinement is strictly necessary or whether the oscillatory terms could alternatively be obtained as the consequence of additional yet unknown action correlations within the double sum over orbits. 
 
\subsection{Generating function}
 
 In this approach, we represent the correlation function through derivatives of a generating function 
\begin{equation}
\label{Zdef} Z(\eps_A,\eps_B,\eps_C,\eps_D)=
\left\langle
\frac{\Delta\left(E+\frac{\eps_C}{2\pi\overline{\rho}}\right)
\Delta\left(E-\frac{\eps_D}{2\pi\overline{\rho}}\right)}
{\Delta\left(E+\frac{\eps_A}{2\pi\overline{\rho}}\right)
\Delta\left(E-\frac{\eps_B}{2\pi\overline{\rho}}\right)}
\right\rangle.
\end{equation}
as
\begin{equation}R(\epsilon)=-2\,{\rm Re}\frac{\partial^2Z(\eps_A,\eps_B,\eps_C,\eps_D)}{\partial\eps_A\partial\eps_B}\big|_{\eps_A=\eps_B=\eps_C=\eps_D=\eps}-\frac{1}{2}
\label{deriv}
\end{equation}
This representation can be derived if we use the relation between the level density and the trace of the Green's function as well as the identity $\det=\exp{\rm Tr}\ln$, ideas that were already combined in (\ref{detint}) \cite{Muller-2009,Haake-2018}. 

We now use the Riemann-Siegel lookalike formula (\ref{rs}) for the two spectral determinants in the numerator. For the spectral determinants in the denominator, no analogous formula is available, and hence we use (\ref{denominator}). 

 There are four possible ways to pair the summands in the two spectral determinants. Using the first summand for $\Delta\left(E+\frac{\eps_C}{2\pi\overline{\rho}}\right)$ and the second summand for $\Delta\left(E-\frac{\eps_D}{2\pi\overline{\rho}}\right)$ leads to 
 a quadruple sum over pseudo-orbits (sets of periodic orbits) $A$, $B$, $C$ and $D$
\begin{align}
\label{Z1} Z^{(1)}\approx& \Bigg\langle \ue^{\pi \ui\left(\bar N \left(E+\frac{\eps_A}{2\pi\bar\rho}\right)-\bar N \left(E-\frac{\eps_B}{2\pi\bar\rho}\right)-\bar N \left(E+\frac{\eps_C}{2\pi\bar\rho}\right)+\bar N \left(E-\frac{\eps_D}{2\pi\bar\rho}\right)\right)}  \sum_{A,B,C,D}F_A F_B^* F_C F_D^*(-1)^{n_C+n_D} 
\ue^{\frac{\ui}{\hbar}\left(S_A \left(E+\frac{\eps_A}{2\pi\bar\rho}\right)-S_B\left(E-\frac{\eps_B}{2\pi\bar\rho}\right)+S_C\left(E+\frac{\eps_C}{2\pi\bar\rho}\right)-S_D\left(E-\frac{\eps_D}{2\pi\bar\rho}\right)\right)}\Bigg\rangle.
\end{align}
Here we have dropped the restriction to cumulative periods smaller than $T_H/2$ from the Riemann-Siegel lookalike formula. In the works cited above, this was justified 
by considering energy increments with small imaginary parts, suppressing the contributions of long cumulative periods, and then analytically continuing towards real energy increments.
 A more recent thorough explanation for dropping this restriction can be found in \cite{Braun-2019}.

 The contribution combining the second summand for $\Delta\left(E+\frac{\eps_C}{2\pi\overline{\rho}}\right)$ and the first summand for $\Delta\left(E-\frac{\eps_D}{2\pi\overline{\rho}}\right)$ can subsequently be obtained by the replacement. 
\begin{equation}
Z^{(2)}(\eps_A,\eps_B,\eps_C,\eps_D)
=Z^{(1)}(\eps_A,\eps_B,-\eps_D,-\eps_C).\end{equation}

Crucially, both $Z^{(1)}$ and $Z^{(2)}$ involve oscillatory factors that contain integrated level densities. However, due to the opposite signs, these oscillations are contained. In contrast, if we pair either the two first summands or the two second summands for both $\Delta\left(E+\frac{\eps_C}{2\pi\overline{\rho}}\right)$ and $\Delta\left(E-\frac{\eps_D}{2\pi\overline{\rho}}\right)$ we obtain rapidly oscillating contributions that average to zero. Hence, the semiclassical approximation of the generating function is given by 
\begin{equation}
\label{Zsum}
Z= Z^{(1)}+Z^{(2)}.
\end{equation}
Taylor expansion using $\frac{d\bar N}{dE}=\bar\rho$ and $\frac{dS}{dE}=T$ now leads to 
\begin{equation}
\label{Z1exp} Z^{(1)}\approx \ue^{\ui(\eps_A+\eps_B-\eps_C-\eps_D)/2} 
\Big\langle \sum_{A,B,C,D}F_A F_B^* F_C F_D^*(-1)^{n_C+n_D} 
\ue^{\ui(S_A(E)-S_B(E)+S_C(E)-S_D(E))/\hbar} \ue^{\ui(T_A\eps_A+T_B\eps_B+T_C\eps_C+T_D\eps_D)/T_H}
\Big\rangle\;.
\end{equation}
 If we now evaluate the correlation function using the prescription (\ref{deriv}), the prefactor in $Z^{(1)}$ turns into $1$. In contrast, the corresponding factor  in $Z^{(2)}$ turns into $\ue^{2\ui\eps}$. Hence $Z^{(2)}$ provides the key to recover oscillatory contributions to the correlation function, which ultimately arise from the exponentiated integrated level densities in the spectral determinants.

Generalizing the previous mechanism,
systematic contributions to the generating functions now arise from quadruplets of pseudo orbits $A,B,C,D$ for which the action difference 
\begin{equation}
\Delta S = (S_A+S_C)-(S_B+S_D)\end{equation}
 is small, i.e., the cumulative action of $B$ and $D$ must be similar to the cumulative action of $A$ and $C$. This happens if the orbits involved in $B$ and $D$ coincide with the orbits involved in $A$ and $C$ except for time reversal (if our system is time-reversal invariant) and/or changing connections inside encounters. When changing connections inside the encounters, also scenarios beyond subsection \ref{sec:sr} have to be taken into account, involving several encounters and encounters with more than two stretches.   
In the following, we denote by $Z_{\rm diag}^{(1)}$ the {\bf diagonal approximation} of $Z^{(1)}$, i.e., its restriction to $A,B,C,D$ where the orbits of $A$ and $C$ are repeated in $B$ and $D$ (modulo time reversal). 
 In contrast, $Z_{\rm off}^{(1)}$ denotes the {\bf off-diagonal part}, where the orbits of $A$ and $C$ are reconnected inside encounters and subsequently included in $B$ and $D$. 
 To account for $A,B,C,D$ where some orbits of $A$ and $C$ are repeated and $B$ and $D$ and others are reconnected, we also have to consider the product of these contributions. Our overall result for $Z^{(1)}$ is therefore 
\begin{equation}
Z^{(1)}=Z_{\rm diag}^{(1)}(1+Z_{\rm
off}^{(1)}).
\end{equation}
To fully appreciate this formula, we note that by convention $Z_{\rm diag}^{(1)}$ includes the case where $A$, $B$, $C$, and $D$ are all empty, which means that no additional summand containing only $Z_{\rm diag}^{(1)}$ is required. The oscillatory factor is included in $Z_{\rm diag}^{(1)}$ only.
 
\subsection{Diagonal part}
 
Let us first evaluate the diagonal part for systems without TRI. For these systems, $Z^{(1)}_{\rm diag}$
 factorizes into sums over the intersection $A\cap B$ (i.e. orbits included in both $A$ and $B$) as well as similar intersections $A\cap D$, $C\cap B$ and $C\cap D$.
The first of these sums is evaluated as 
\begin{equation}
\sum_{A\cap B} |F_{A\cap B}|^2\ue^{\ui T_{A\cap B}(\eps_A+\eps_B)}
=\exp\left(\sum_p |F_p|^2 \ue^{\ui T_p(\eps_A+\eps_B)}\right)
=\exp\left(\int_0^\infty \frac{\ud T}{T}\ue^{\ui T(\eps_A+\eps_B)}\right)
=\exp\left(-\ln(\eps_A+\eps_B)\right)=\frac{1}{\eps_A+\eps_B}.\end{equation}
Here, we have used that the sum over pseudo-orbits with $m$ members can be written as the $m$-fold power of a single orbit sum, divided by $m!$ to account for different orderings of these members. 
Thus, it agrees with the Taylor expansion of an exponentiated orbit sum
and can be evaluated using the sum rule (\ref{sumrule}). The sum over $B\cap D$ is analogous. In contrast, for $A\cap D$ and $C\cap B$ the sign factor $(-1)^{n_c+n_D}$ turns into $(-1)^m$ and the result is inverted. For time-reversal invariant systems, doubling the orbit sum is doubled, hence the result is squared.\footnote{Cases where an orbit is included in one of the pseudo-orbits multiple times, defying the set notation employed here, have a negligible impact on the result for reasons similar to the previous diagonal approximation.} 
Thus, we obtain the overall result
\begin{equation}
\label{Z1diag}
Z_{\rm diag}^{(1)}=\ue^{\ui\left(
\eps _{A}+\eps _{B}-\eps _{C}-\eps _{D}\right)/2 }%
\left(
\frac{\left( \eps _{A}+\eps _{D}\right) \left( \eps
_{C}+\eps _{B}\right) }{\left( \eps _{A}+\eps
_{B}\right) \left( \eps _{C}+\eps _{D}\right) }\right)^\kappa.\end{equation}
In the literature, analogous steps were related to a cumulant expansion in \cite{Heusler-2007, Muller-2009}, and 
to a dynamical zeta function in \cite{Keating-2007}.

For {\bf systems without TRI}, applying the prescription (\ref{deriv}) to the diagonal approximation alone already reproduces the GUE result. This is similar to the approach based on a double sum over periodic orbits, where the diagonal approximation gave the full non-oscillatory part of the GUE result. Hence in this case, while encounter contributions are present, we expect them to eventually cancel. In contrast, for time-reversal invariant systems, encounters should give important non-canceling contributions.

  \subsection{Off-diagonal part}

 We now proceed to evaluate the off-diagonal part. To do so, we extend the definition of an encounter and speak of an {\bf $l$-encounter} whenever $l$ stretches of one periodic orbit or several orbits come close. Again, this proximity may be modulo time reversal if our system is time-reversal invariant. We can again change connections inside the encounter, relying on an argument with stable and unstable directions similar to subsection \ref{sec:sr}. Changing connections also modifies the links in between the encounters, but to a much lower degree. Changing connections may leave the number of orbits unchanged as before, but it also may split periodic orbits into separate components, or merge orbits with each other.
 
\begin{figure}[h]
\centering
\includegraphics[width=0.55\textwidth]{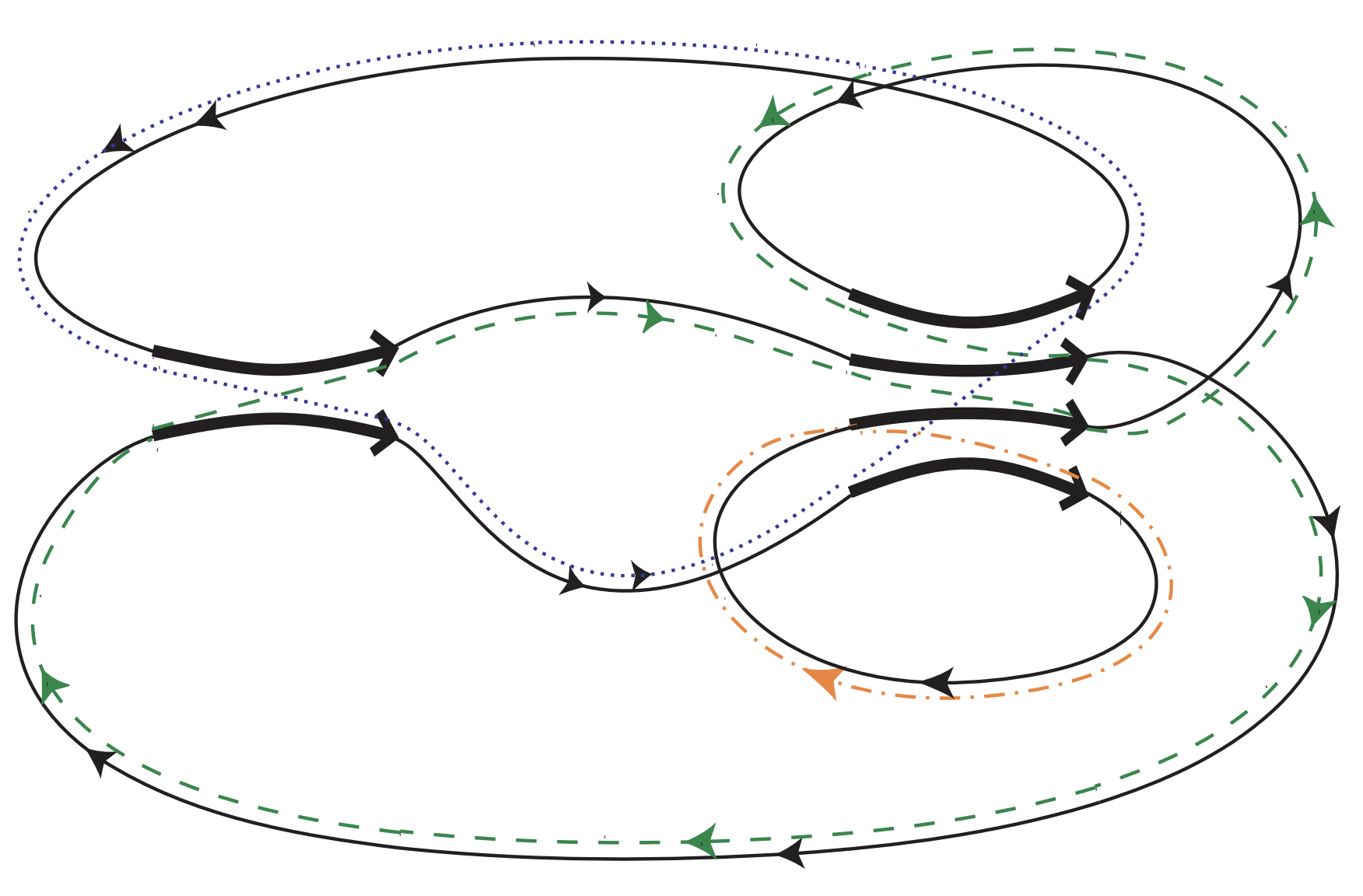}
\caption{Diagram involving one 2-encounter and one 4-encounter. Upon changing connections, the black orbit decomposes into three orbits in blue, green and orange.}
\label{24}
\end{figure}

\begin{figure}[h]
\centering
\includegraphics[width=0.55\textwidth]{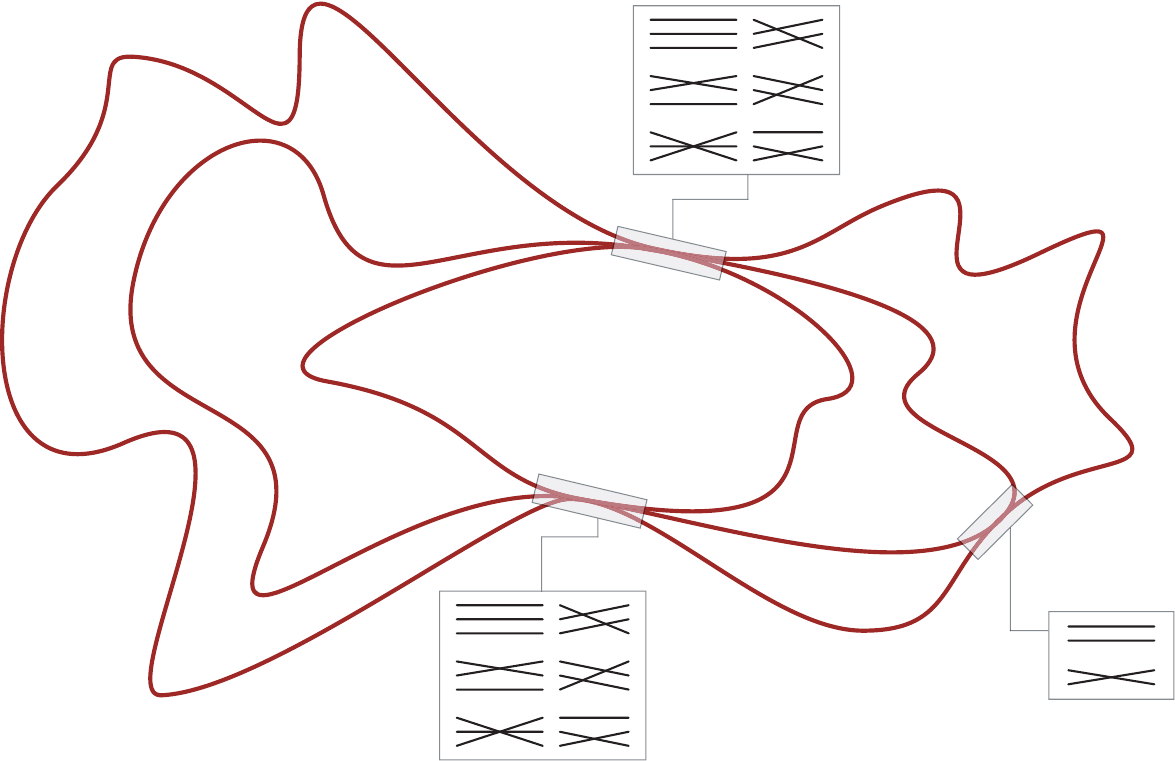}
\caption{Example for a bunch of orbits.}
\label{bunch}
\end{figure}

  An example is shown in Fig. \ref{24}, where one orbit (depicted in black) contains one 2-encounter and one 4-encounter. After changing connections, this orbit decomposes into three separate orbits depicted in blue, green, and orange. The stretches in both encounters are almost parallel and close in phase space, as opposed to almost mutually time-reversed. Hence correlated orbits as sketched in this figure arise even if the system is not time-reversal invariant.
 Similarly, in Fig. \ref{bunch}  any of the depicted ways of choosing connections in two 3-encounters and one 2-encounter lead to either one permissible orbit or a   pseudo-orbit decomposing into several orbits. In general, periodic orbits (and pseudo-orbits) form {\bf bunches} \cite{Altland-2008} whose members differ by their connections inside encounters.
 
\begin{figure}[t]
\centering
\includegraphics[width=0.99\textwidth]{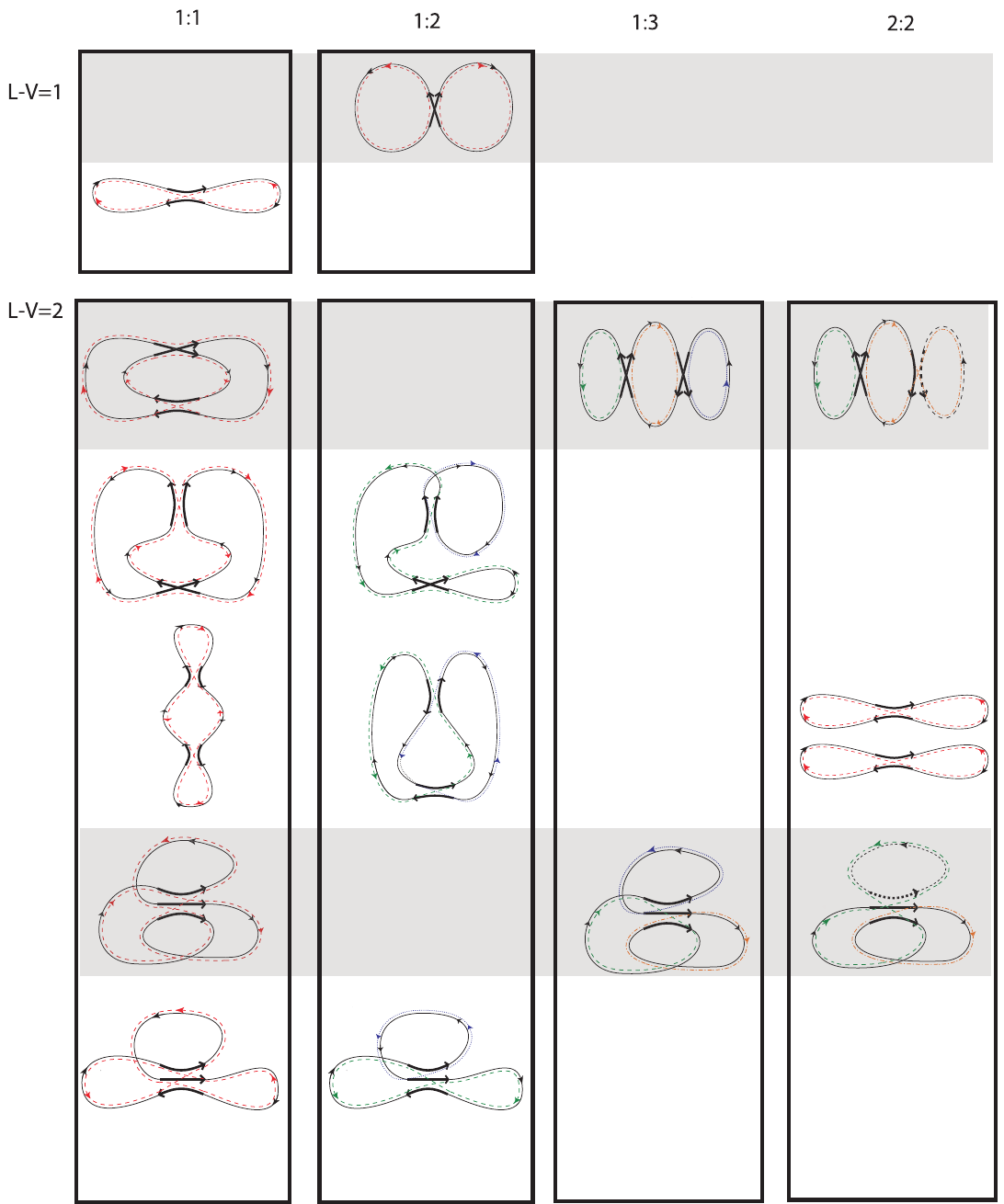}
\caption{All diagrams contributing to the form factor with $L-V=1$ or $L-V=2$. The headings $j:k$ give the number of original and partner orbits. Diagrams not requiring time-reversal invariance are highlighted in gray.}
\label{leading}
\end{figure}

 A fuller collection of examples, to be referred to again later, is shown in Fig. \ref{leading}. The diagrams in \ref{leading} with a gray background do not require time reversal invariance, as the stretches involved in the encounters point in parallel directions. In contrast, the remaining diagrams involve at least some stretches that are approximately mutually time-reversed, and hence require TRI.

 To obtain systematic contributions to the generating function, the pseudo-orbits $B$ and $D$ must have a similar cumulative action to the pseudo-orbits $A$ and $C$. Hence, the diagrams depicted will contribute if the orbits before reconnection are distributed in some way between $A$ and $C$ whereas the orbits after reconnection are distributed between $B$ and $D$. 

Following (\ref{Z1exp}) the contribution of each such class of quadruplets to the generating function involves a sign factor $(-1)^{n_C+n_D}$.
It also involves factors arising from each link and from each encounter. We refer to \cite{Muller-2009,Haake-2018} for the derivation of these factors. However, the results are in close analogy to subsection \ref{sec:srcontrib}.  The analogy becomes even greater noting that we could have left  $2\eps$ as $\eps_1+\eps_2$ in (\ref{doublesum}), with $\eps_1$ corresponding to the original orbit $p$ and $\eps_2$ corresponding to the partner orbit $p'$.  

Each {\bf link} belongs to one orbit before reconnection and this orbit may be part of either pseudo-orbit $A$ or $C$. Likewise, it belongs to one of the pseudo-orbits $B$ or $D$ after reconnection. Hence, the natural generalization of the link factor $-\frac{1}{\ui(\eps_1+\eps_2)}$ for this case is 
\begin{equation}
-\frac{1}{\ui(\eps_{\text{$A$ or $C$}}+\eps_{\text{$B$ or $D$}})}.
\end{equation}

Similarly, the $l$ stretches of an {\bf $l$-encounter} before reconnection must be distributed in some way among the pseudo-orbits $A$ and $C$. We denote the corresponding numbers of stretches by $l_A$ and $l_C$ with $l=l_A+l_C$. After reconnecting, we have the corresponding numbers $l_B$ and $l_D$ with $l=l_B+l_D$. The encounter factor $2\ui(\eps_1+\eps_2)$ can be generalized to 
\begin{equation}
\ui(l_A\eps_A+l_B\eps_B+l_C\eps_C+l_D\eps_D).
\end{equation}

\subsection{Combinatorics}

\begin{figure}[h]
\centering
\includegraphics[width=.3\textwidth]{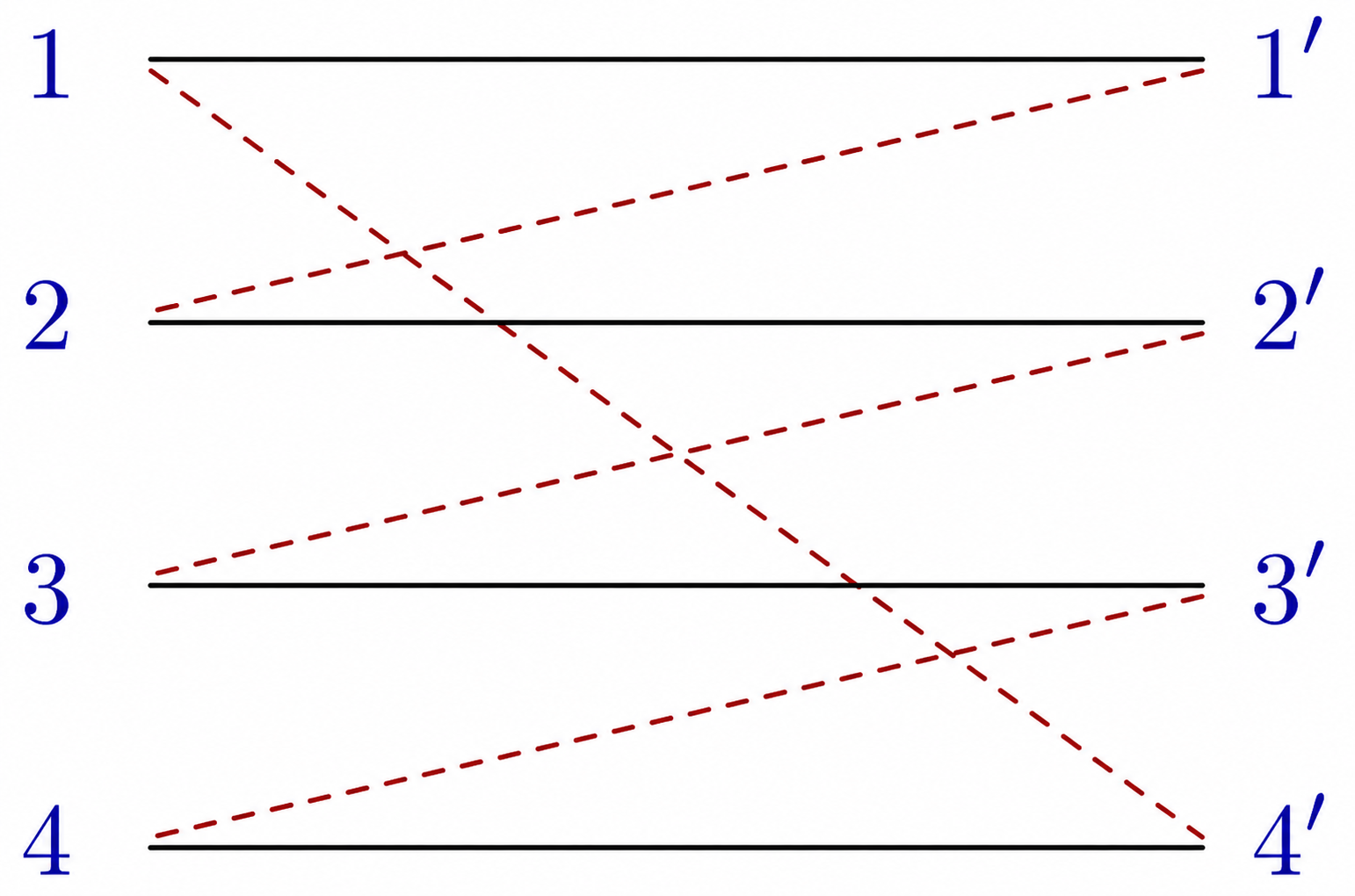}
\caption{Convention for the numbering of stretches. The stretches before reconnection are shown in black. After reconnection, the entrance port of the $j$th stretch is connected to the exit port of the $(j-1)$st stretch, and the entrance port of the first stretch is connected to the exit port of the final stretch, as illustrated by the dashed red line.}
\label{4enc}
\end{figure}
 
To show agreement with RMT, we have to sum over infinitely many classes of quadruplets of pseudo orbits. This requires a clean and convenient way to characterize these classes. We  hence sketch the definition of {\bf structures} of quadruplets from \cite{Muller-2009}. In this definition, the encounters are numbered in an arbitrary way. Afterwards, the stretches involved in each encounter are numbered in such a way that after reconnection, the beginning of the stretch $j$ (called the {\bf entrance port}) is connected to the end ({\bf exit port})  of stretch $j-1$, as depicted in Fig. \ref{4enc}. (The beginning of the final stretch is then connected to the end of the first stretch.) Now, a structure is characterized by 
\begin{itemize} 
\item  numbers of encounters and stretches,
\item  a way in which the ends of encounter stretches are connected through links,
\item  a way to distribute the periodic orbits before reconnection among $A$ and $C$ and a way to distribute the orbits after reconnection among $B$ and $D$.
\end{itemize} 
For systems without time reversal invariance, all stretches of an encounter must be traversed with the same direction of motion, and hence links must connect opposite sides of the encounters. For time-reversal invariant systems, these restrictions are no longer in place, and moreover the directions of the reconnected orbits may be chosen freely. (We note that in this case, our convention requires connections as in Fig. \ref{4enc} regardless of the directions of the stretches.) 

Most diagrams of quadruplets, such as depicted in Figs. \ref{24}, \ref{bunch}, and \ref{leading} can be described by several {\bf equivalent structures}, differing in the numbering of encounters and stretches. The number of these equivalent structures depends on the symmetries of the diagram. (Because
as for a symmetrical diagram, some ways of changing the numbering of encounters and stretches may lead to the same structure, whereas for a diagram without any symmetries, all changes of numbering lead to different structures.)
This situation is loosely reminiscent of the multiplicities of Feynman diagrams in perturbation theory.

As seen in \cite{Heusler-2007,Muller-2009}, it is cleanest to write $Z^{(1)}$ as a sum over structures.  This also entails a change of the {\bf encounter factor} from $\ui(l_A\eps_A+l_B\eps_B+l_C\eps_C+l_D\eps_D)$ to 
\begin{equation}
\label{encnew}
\ui(\eps_{A\;{\rm or}\;C}
+\eps_{B\;{\rm or}\;D}),
\end{equation}
which is even more similar to the link factor. In the new encounter factor, the choice of $A$ versus $C$ and $B$ versus $D$ is taken based on the pseudo-orbits to which the beginning of the first stretch of the encounter belongs.  The previous factor is then recovered by summation over equivalent structures. 

Altogether, we then obtain the following expression for the {\bf off-diagonal part of the generating function}:
\begin{equation}
Z^{(1)}_{\rm off}=\sum_{\rm structures} \frac{1}{V!}(-1)^{n_C+n_D} \frac{\prod_{\rm
enc}\ui(\eps_{A\;{\rm or}\;C}
                           +\eps_{B\;{\rm or}\;D})}
     {\prod_{\rm links}(-\ui(\eps_{A\;{\rm or}\;C}
                           +\eps_{B\;{\rm or}\;D}))}\,
\end{equation}

If we denote the number of encounters by $V$ and the number of links (equal to the number of encounter stretches) by $L$, we see that the contribution of each structure to the generating function is of the order $\frac{1}{\eps^{L-V}}$. Upon taking derivatives as in (\ref{deriv}) we thus expect a contribution to the correlation function of order $\frac{1}{\eps^{L-V+2}}$.\footnote{We note that this conclusion is not immediate if the derivative acts on the oscillatory factor.} 

Hence, the diagrams with $L-V=1,2$ shown in Fig. \ref{leading} are responsible for orders $\frac{1}{\epsilon^3}$ and $\frac{1}{\epsilon^4}$. The diagrams contributing to $\frac{1}{\epsilon^3}$ involve a {\bf single 2-encounter} ($L=2$, $V=1$). The case of two almost time-reversed stretches was considered before. If the stretches are approximately parallel, the orbit decomposes upon reconnection, and contributions are obtained only in the approach involving quadruplets of pseudo-orbits. However, these contributions cancel if we sum over different assignments of the orbits to $A,B,C$ and $D$ \cite{Muller-2009}. This is anticipated, as the diagram does not require TRI, and for systems without TRI encounter contributions should cancel.

The diagrams contributing to $\frac{1}{\epsilon^4}$ involve either {\bf two 2-encounters} ($L=4$, $V=2$) or {\bf one 3-encounter} ($L=3$, $V=1$). Those in the column '1:1' were identified as responsible for the non-oscillatory $\frac{1}{\epsilon^4}$ contribution in \cite{Heusler-2004} (corresponding to $\tau^3$ in the form factor), with a cancelation between the two diagrams not requiring TRI. The other diagrams have to be considered in the generating function approach as well, allowing to  recover the oscillatory term proportional to $\frac{1}{\epsilon^4}\ue^{2\ui\eps}$.

\begin{BoxTypeA}

\noindent 
{\bf Exercise:} Determine the numbers of equivalent structures for each of the diagrams in the 1:1 column of Fig. \ref{leading}, assuming one fixed assignment to pseudo-orbits (e.g. that the original orbit is included in $A$ and the partner orbit in $B$).

\end{BoxTypeA}

\begin{BoxTypeA}

\noindent 
{\bf Exercise:} Which orders in $\frac{1}{\epsilon}$ does the bunch in Fig. \ref{bunch} contribute to? What does it mean if the connections inside an encounter cannot be brought to a form similar to Fig. \ref{4enc}?

\end{BoxTypeA}

For {\bf systems without TRI}, the cancelation of encounter contributions could be demonstrated in \cite{Muller-2009} for the full generating function. The essential idea was to find a mapping between structures that cancel pairwise. One of the canceling structures involves a 2-encounter, an arbitrary $l$-encounter, and possibly more encounters. In the other, one stretch of the 2-encounter and one stretch of the $l$-encounter are merged, forming an $(l+1)$-encounter.

For {\bf systems with TRI} agreement of the semiclassical result with the GOE could be demonstrated in full only by establishing a connection between the semiclassical approach and the sigma model, which we will turn to later. However, for the case of non-oscillatory contributions, a direct derivation is also possible \cite{Muller-2004,Muller-2005}. This involves establishing a recursion between subsequent orders in $\frac{1}{\eps}$, or equivalently, $\tau$.  An essential step in deriving this recursion is to relate structures contributing to different orders, specifically structures where the number of stretches in a given encounter requiring TRI differ by one. 

To implement this recursion, it is helpful to describe structures in terms of {\bf permutations} i.e., bijective mappings between integers. We need three permutations: one that describes how the beginnings of the encounter stretches follow each other along the original orbit(s), one that describes how these follow along the partner orbit(s), and one that describes how connections are changed in the encounters. We can then write the partner orbit permutation as a product of the original orbit permutation and the encounter permutation. Cycles of the original orbit and partner orbit permutations correspond to periodic orbits, whereas cycles of the encounter permutation correspond to encounters. We refer to \cite{Muller-2005} (see also the arxiv version of \cite{Muller-2009}) for details of this approach. The description of structures in terms of permutations is also helpful in order to systematically search for relevant diagrams in related problems, such as for higher-order correlation functions in \cite{Mueller-2018a}.

\subsection{Relation to the sigma model}

However, a full semiclassical derivation of the RMT results, including both oscillatory contributions and the orthogonal case, currently requires 
establishing a connection to the sigma model. We will give an overview of the essential ideas referring to \cite{Muller-2009} (as well as \cite{Muller-2005,Heusler-2007}) for the original articles and \cite{Haake-2018} for a detailed pedagogical account.  For didactical reasons, our presentation will focus on the unitary case initially, 
skip over delicate issues associated to the treatment of signs,
and it will be phrased as a comparison between semiclassics and the sigma model. However, we stress that the approach taken has the status of a semiclassical {\bf derivation} of a sigma model for both the unitary and the orthogonal case. The sigma model thus derived coincides with the one from RMT, i.e., it does not include non-universal corrections that alternative approaches (to be discussed in section \ref{sec:outlook}) have aimed to recover.

The sigma model also recovers the correlation function via the generating function (\ref{Zdef}). In its RMT version, the individual quantum Hamiltonian is replaced by a large matrix averaged over with a Gaussian weight. It is then convenient to express the determinants in the denominator through Gaussian integrals over complex numbers. Similarly, the determinants in the numerator can be represented through Gaussian integrals over anticommuting (Grassmannian) variables. Given the analogy to anticommuting creation and annihilation operators, these variables are also referred to as fermionic, forming a counterpart of complex bosonic variables. Both sets of variables can be treated on equal footing, in the spirit of a {\bf supersymmetric} approach. Hence the ratio of four determinants is written as an integral over supervectors with bosonic and  fermionic components. The approach also involves supermatrices in $2\times 2$ block form, with bosonic diagonal blocks and fermionic off-diagonal blocks. 

An alternative representation for the ratio of four determinants can be based on the replica trick,   representing the determinants in the numerator as $(r-1)$ times the powers of inverse determinants and then taking the limit $r\to0$. The replica approach was connected to semiclassics in appendices of the arxiv version of \cite{Muller-2009} and in \cite{Muller-2005}.

In the {\bf derivation of the supersymmetric sigma model from RMT}, one first evaluates the average over Hamiltonian matrices for fixed supervectors. Then the remaining supervector integral is replaced by a supermatrix integral (removing an exponentiated term quartic in the former) using a Hubbard-Stratonovich approximation. The supermatrix integral is subsequently evaluated in a saddle-point approximation. There are two saddles, termed the standard and the Andreev-Altshuler saddle point, which correspond to our $Z^{(1)}$ and $Z^{(2)}$. Explicit formulas for their contribution can be written in the so-called rational parameterization, which is not necessarily the first choice for work in RMT but has the advantage of a close connections to semiclassics.
The expression for $Z^{(1)}$ arising after all these steps is
\begin{equation}
Z^{(1)}=-\ue^{\ui\,(\eps_A+\eps_B-\eps_C-\eps_D)/2}
\int \ud[B,\tilde B]
\exp\left(\ui\,{\rm Str}\,\sum_{l=1}^\infty
\left(\hat\eps(\tilde{B}B)^l+\hat\eps'(B\tilde{B})^l\right)
\right).
\end{equation}
with the same connection between $Z^{(1)}$ and $Z^{(2)}$ as in semiclassics. For the unitary case,
$B$ is a supermatrix given by $B=\begin{pmatrix}B_{11}&B_{12}\\B_{21}&B_{22}\end{pmatrix}$ and $\tilde B=\begin{pmatrix}B^*_{11}&B_{21}^*\\B_{12}^*&-B_{22}^*\end{pmatrix}$ is its adjoint apart from a sign factor. 
${\rm Str}$ indicates the supertrace, i.e. for a $2\times 2$ supermatrix the difference between the diagonal elements, and we have $\hat\eps={\rm diag}(\eps_1,\eps_2)={\rm diag}(\eps_A,\eps_C)$, 
$\hat\eps'={\rm diag}(\hat\eps_1,\hat\eps_2)={\rm diag}(\eps_B,\eps_D)$. 
If we restrict ourselves to the summand $l=1$, we obtain the Gaussian (or more precisely Fresnel) integral
\begin{equation} 
\label{sigmadiag}
-\int \ud[B,\tilde B]
\exp\left[\ui\,{\rm Str} 
\left(\hat\eps\tilde{B}B+\hat\eps'B\tilde{B}\right)
\right]
=\frac{\left( \eps _{A}+\eps _{D}\right) \left( \eps
_{C}+\eps _{B}\right) }{\left( \eps _{A}+\eps
_{B}\right) \left( \eps _{C}+\eps _{D}\right) },
\end{equation}
coinciding with the diagonal approximation. 
The derivation uses that the exponent can be written in terms of $\ui(\epsilon_j+\hat\epsilon_k)B_{kj}{\tilde B}_{jk}$.
Defining an average $\left(...\right)$ with a normalized Gaussian  weight, we can thus rewrite our full expression for $Z^{(1)}$ as 
\begin{equation}
Z^{(1)}=\ue^{\ui\,(\eps_A+\eps_B-\eps_C-\eps_D)/2}\frac{\left( \eps _{A}+\eps _{D}\right) \left( \eps
_{C}+\eps _{B}\right) }{\left( \eps _{A}+\eps
_{B}\right) \left( \eps _{C}+\eps _{D}\right) }
\left\langle
\exp\left(\ui\,{\rm Str}\,\sum_{l=2}^\infty
\left(\hat\eps(\tilde{B}B)^l+\hat\eps'(B\tilde{B})^l\right)
\right)\right\rangle
\end{equation}
Now, the exponential can be expanded into a Taylor series. We are thus led to an average over products of supertraces, which can also be written out in components. For the example of $l=2$, this involves summands
\begin{equation}
\label{sigma2}
\ui(\eps_{j_1}+\hat\eps_{k_1})B_{k_1j_1}{\tilde B}_{j_1k_2}B_{k_2j_2}{\tilde B}_{j_2k_1}, \end{equation}
and analogous formulas hold for larger $l$.
Applying Wick's 
 theorem (for complex and Grassmannian variables) to the Gaussian average, we then obtain a sum over contractions between matrix elements of $B$ and $\tilde B$ that are mutually complex conjugated.  
 
 \begin{figure}[h]
\centering
\includegraphics[width=.4\textwidth]{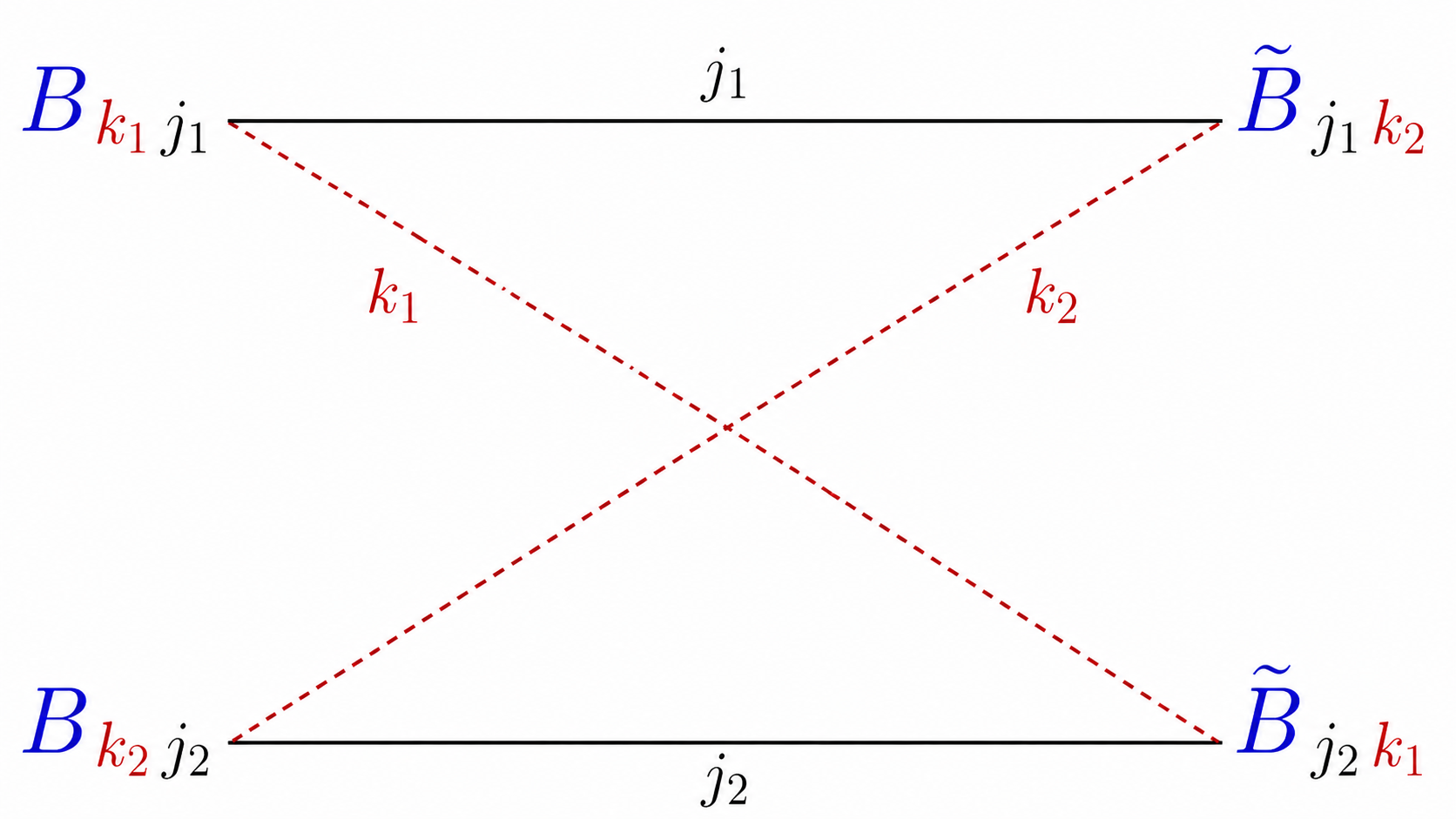}
\caption{Correspondence to the sigma model for entrance and exit ports for a 2-encounter.
The indices $j_1,j_2$ correspond to pseudo-orbits before reconnection, and the indices $k_1,k_2$ correspond to pseudo-orbits after reconnection.}
\label{indices}
\end{figure}
 
 We are now prepared to demonstrate the {\bf correspondence to encounter diagrams} arising in semiclassics.
 Each supertrace factor corresponds to an {\bf encounter} with $l$ stretches, and the matrix elements of $B$ and $\tilde B$ correspond to the endpoints (ports) of the  stretches on the two sides of the encounter. We can interpret the indices of these matrix elements as indicating the pseudo-orbits to which the corresponding port belongs. The indices $j$ indicate the original pseudo-orbits with $j=1$  (bosonic) corresponding to $A$ and $j=2$ (fermionic) corresponding to $C$. The index $k$ indicates the partner pseudo-orbit with $k=1$ (bosonic) corresponding to $B$, and $k=2$ (fermionic) corresponding to $D$. The consistency of this interpretation can be checked as follows: As every entrance port (matrix element of $B$) belongs to the same original orbit and hence pseudo-orbit as the following exit port (matrix element of $\tilde B$), the two indices $j$ should coincide. 
 This constraint is automatically satisfied due to the nature of the matrix product. Similarly, following the convention of Fig. \ref{4enc} each entrance port belongs to the same partner pseudo-orbit as the preceding exit port and hence shares its index $k$, which is again guaranteed by the form of the matrix product. 
 For the case $l=2$, this correspondence between indices is illustrated in Fig. \ref{indices}
 Eq. (\ref{sigma2}) also illustrates that we obtain a factor
 \begin{equation}
\ui(\eps_{j_1}+\hat\eps_{k_1})=\ui(\eps_{A\;{\rm or}\;C}
+\eps_{B\;{\rm or}\;D})
\end{equation}
from every encounter, depending on the pseudo-orbit memberships of its first entrance port. This is in line with the semiclassical result in (\ref{encnew}).

Finally,  contraction lines should be associated to ${\bf links}$. Like links, these lines connect entrance ports (associated with $B$) and exit ports (associated with $B^*$) with identical indices, and hence membership in the same original and partner-pseudo orbits. Given the Gaussian weight above, Wick's theorem  affords the desired link factor  
\begin{equation}
-\frac{1}{\ui(\eps_j+\hat\eps_k)}=-\frac{1}{\ui(\eps_{A\;{\rm or}\;C}
+\eps_{B\;{\rm or}\;D})}.
\end{equation}

For {\bf systems with TRI}, the structure of $B$ and $\tilde B$ becomes more complicated. In the semiclassical picture, their subscripts should be interpreted as expressing both pseudo-orbit membership and direction of motion. 
This form also means that some entries of $B$ are complex conjugated with respect to each other, and the same holds for $\tilde B$. Hence, also contractions (and hence links) involving two ports on the same side of the encounter are possible, as expected based on the relevant semiclassical diagrams. 
The ratio of sums arising 
from the Gaussian integral (\ref{sigmadiag})
is also squared, mirroring the result of the 
diagonal approximation is squared as expected.
Hence, we again obtain full agreement between semiclassics and RMT.


\section{Overview of related work}

\label{sec:outlook}

We conclude with an overview of related topics in the literature.
The semiclassical approach to spectra discussed here has close connections to semiclassical work on {\bf transport} as reviewed in \cite{Novaes-2026}. Omitting work on transport and focusing on spectral problems, recent themes include the following.

\subsection*{Higher-order correlations}

The original diagonal approximation was generalized to {\bf higher-order correlation functions} in \cite{Shukla-1997}. For systems without TRI, this was extended to include oscillatory contributions in \cite{Nagao-2009}. For non-oscillatory contributions, the leading encounter diagrams were identified in \cite{Mueller-2018a}, and all diagrams were shown to sum to zero in absence of TRI in \cite{Mueller-2018b}.

\subsection*{Other systems}

A trace formula for systems with {\bf spin} was derived in \cite{Bolte-1998} and applied to spectral statistics in \cite{Bolte-1999,Bolte-2006,Braun-2012}.

{\bf Quantum graphs} are reviewed in this volume in \cite{Berkolaiko-2026}. An exact trace formula for quantum graphs was derived in \cite{Kottos-1999} and spectral statistics was studied using methods based on the diagonal approximation and encounters in \cite{Kottos-1999,Berkolaiko-2003,Berkolaiko-2004}. In an alternative approach to spectral statistics, a sigma model for quantum graphs was justified in \cite{Gnutzmann-2004, Gnutzmann-2005, Pluhar-2014, Pluhar-2015}.

Recently, there has been a lot of interest in {\bf quantum chaotic many-body systems}. In this volume, a general introduction to this topic is given in \cite{Urbina-2026}, with a review of periodic-orbit approaches in \cite{Waltner-2026}. 
A key idea behind this work is that replacing creation and annihilation operators in a second quantized Hamiltonian by complex variables (with the interpretation of a macroscopic wave function) shares many features with the traditional classical limit of a general quantum system. Developments close to the material reviewed in the present chapter include an extension of trace formula methods in \cite{Engl-2015,Dubertrand-2016}, their application to spectral statistics in \cite{Dubertrand-2016}, as well as the identification of further action correlations substantially extending the encounter mechanism in \cite{Gutkin-2016}. 

An important model system for many body quantum mechanics are {\bf quantum circuits}, reviewed in \cite{Prosen-2026}. An approach to spectral statistics for these circuits was developed in \cite{Bertini-2021}.

\subsection*{Symmetries}

Methods from periodic-orbit theory have been extended to systems with discrete and continuous {\bf symmetries} in \cite{Robbins-1989,Creagh-1990,Keating-1997,Joyner-2012,Blatzios-2025}. 
For the spectral statistics of these systems, the methods reviewed in this chapter cannot be applied without adaptations, as there are additional correlations between orbits related by symmetry operations. 
In the discrete case, this challenge could be addressed by using trace formula adapted to subspectra that can be associated to irreducible representations of the symmetry group. 
Interestingly, the spectral statistics in the subspectra can differ from expectations based on the time-reversal properties of the full system. 

For systems with {\bf arithmetic chaos}, specific symmetries entirely change the nature of spectral correlations \cite{Bogomolny-1997}. A key example are certain billiards (e.g. triangles with specific angles) on hyperbolic surfaces. These systems are fully chaotic, but display an infinite number of so-called {\bf Hecke symmetries} commuting with the Hamiltonian. These symmetries turn the level statistics into Poissonian form. Arithmetically chaotic systems allow only for discrete periodic orbit lengths (and hence actions), leading to large action degeneracies that completely change the semiclassical approach. However, a change of boundary conditions may lead to pseudo-arithmetic chaos, where the Maslov indices cause a cancelation of the additional contributions from action-degenerate orbits, and hence RMT statistics prevail \cite{Braun-2010,Braun-2015}.
 
The threefold classification of symmetry classes leading to the GUE, GOE and GSE was extended by Altland and Zirnbauer \cite{Altland-1997} to a {\bf tenfold} classification, involving antiunitary operators commuting with the Hamiltonian (generalized time-reversal symmetries), antiunitary operators anticommuting with the Hamiltonian, and unitary operators anticommuting with the Hamiltonian. For certain quantum graphs that display these (pseudo) symmetries, trace formulas and a semiclassical approach to spectral statistics were developed in \cite{Gnutzmann-2003,Gnutzmann-2004a}. A key case are {\bf graph models of Andreev billiards}. An Andreev billiard consists of a normal conductor with a boundary consisting of an insulator and a superconductor. Reflection at the superconductor is self-retracing and converts electrons into holes and vice versa. A suitable generalization of the diagonal approximation thus involves self-dual orbits, where the actions accumulated as an electron and as a hole exactly cancel. These orbits then lead to systematic deviations of the average level density from the Weyl term, which reproduce (partly or fully) predictions from random-matrix ensembles associated to the tenfold way. 

\subsection*{Andreev gap}

On a related note, a gap in the spectrum of Andreev billiards was explained by Kuipers et al in \cite{Kuipers-2010,Kuipers-2011}.  The authors used a semiclassical approach that is not directly based on a trace formula but on an expression of the level density in terms of a trace of products of scattering matrices. Using a semiclassical expression for these scattering matrix elements, they obtained a multiple sum over trajectories, and studied the case where these trajectories coincide pairwise or are related by changing connections in encounters. They then solved the combinatorial problem of summing over all relevant diagrams by introducing a generating function and showing that it satisfies an algebraic equation. Studying the solutions of this equation ultimately leads to an explanation of the gap. 

\subsection*{Rigorous work}

The semiclassical approach to spectral statistics  can be made more rigorous than in the present account. In particular,  the equidistribution theorem \cite{Sieber-2001,Turek-2005} and mixing \cite{Muller-2005} can be used to study the statistics of encounters, and we have already mentioned further work on generating functions \cite{Braun-2019}. However, the existing work does not constitute a mathematical proof. A promising starting point for rigorous mathematical work in this direction is {\bf Selberg's trace formula} \cite{Selberg-1956} for compact Riemannian surfaces, which had already been used in \cite{Sieber-2001}.  For the Selberg trace formula, steps towards a rigorous justification of the semiclassical approach to spectral statistics were presented in \cite{Huynh-2015}.

A related, but conceptually different rigorous approach to spectral statistics was recently developed by Rudnick \cite{Rudnick-2023} for the GOE and by Marklof and Monk \cite{Marklof-2024} for the GUE and GSE. This approach applies to averages over ensembles ({\bf moduli spaces}) of surfaces in the limit of high genus, which are larger than an individual system but smaller than a random-matrix ensemble. The moduli spaces can be adapted to desired time-reversal properties.
The authors showed that the spectral number variance of the Laplacian, when averaged over these moduli spaces, agrees with the corresponding RMT ensemble. The proofs make use of Mirzakhani's theory of moduli spaces and contain elements loosely analogous to a diagonal approximation.

\subsection*{Ballistic sigma model}

In the present account, we have used connections to the sigma model of RMT to help sum over all semiclassical contributions. However, one might ask whether a more immediate connection to a sigma model can be established. Indeed, work in the 1990s \cite{Muzykantskii-1995,Andreev-1996} already aimed at establishing a so-called ballistic sigma model to address universal statistics, with an averaging analogous to RMT averages or disorder averages provided by an energy average, and then recovering RMT as a limit. Although there has been substantial debate about open questions in this early work, its connection to semiclassics has been studied in \cite{JMuller-2007}. 

Recent work on a more direct justification of a sigma model has already been mentioned in the context of quantum graphs. Key further references are \cite{Altland-2015} and chapter 7 of \cite{Haake-2018}.

\subsection*{Beyond universality}
 A key question is what happens to semiclassics and spectral statistics when the conditions for agreement with RMT are not all met.

Outside the strict semiclassical limit, there can be corrections from {\bf diffractive orbits}; see \cite{Bogomolny-2000,Sieber-2000} and references therein. Another important source of corrections are terms proportional to the ratio of {\bf Ehrenfest time} (logarithmic in $\hbar^{-1}$) and Heisenberg time \cite{Tian-2004,Brouwer-2006,Waltner-2010}. 
Although Ehrenfest corrections are outside the strict semiclassical limit for spectral statistics, the situation is more subtle for quantum transport.  Ehrenfest corrections have often 
required a treatment of subtle issues such as `fringes' of an encounter where only some of the stretches remain close.
 In the aforementioned work on Andreev gaps \cite{Kuipers-2010,Kuipers-2011}, Ehrenfest time effects both modify the gap and lead to the prediction of a second gap.  
 
For systems with {\bf mixed phase space}, bifurcations present important challenges \cite{Schomerus-1997}, but work in the diagonal approximation was reported in \cite{Gutierrez-2007}. 

For many-body systems, {\bf collective modes} \cite{Akila-2017} are expected to impact spectral properties.

A potential source of corrections to universal spectral statistics is also given by any {\bf action correlations} that go beyond the encounter diagrams and their natural extensions.

An example for systems that are chaotic but do not satisfy the full conditions from section 2 are {\bf quasi one-dimensional systems} displaying Anderson localization. These include long wires with chaos induced by an irregular shape or impurities. Semiclassical work to investigate the spectral statistics of these systems can be found in \cite{Dittrich-1996,Braun-2008}. For  arrays of chaotic cavities, a mapping to the Dorokhov-Mello-Pereyra-Kumar (DMPK) equation, a hallmark of Anderson localization, was achieved in \cite{Brouwer-2008}.
  
\begin{ack}[Acknowledgments]

We would like to thank our collaborators. In particular, this chapter is closely connected to joint work with Alexander Altland, Petr Braun, Fritz Haake, Stefan Heusler, Jon Keating and Klaus Richter.
	
\end{ack}

\bibliographystyle{JHEP}%

\bibliography{PeriodicOrbits.bib}

\end{document}